\documentclass[11pt]{article}
\usepackage{epstopdf}                                                          
\usepackage[a4paper,hmarginratio=1:1,vmarginratio=2:3,
  totalwidth=15.28cm,totalheight=22.65cm]{geometry}
\usepackage{bm,epstopdf,epsfig,amsmath,amssymb,
amsfonts,colordvi,wrapfig,comment,cancel,verbatim,slashed}
\usepackage{graphicx,graphics}
\usepackage[font=md,captionskip=8pt]{subfig}
\usepackage[usenames,dvipsnames]{color}
\usepackage[noadjust]{cite}
\usepackage{xcolor} 
\usepackage[utf8]{inputenc}
\usepackage{setspace}

\usepackage[utf8]{inputenc}

\newcommand{\sg}{\sqrt{g}}    
\newcommand{\w}{\omega}  

\newcommand{\tGamma}{\tilde\Gamma}
\allowdisplaybreaks
\newcommand{\q}{\alpha}

\newcommand{\cA}{{\cal A}}

\newcommand{\cF}{{\cal F}}

\newcommand{\cI}{{\cal I}}

\newcommand{\cO}{{\cal O}}

\newcommand{\tr}{\text{tr}}

\newcommand{\be}{\begin{equation}}
\newcommand{\ee}{\end{equation}}
\newcommand{\bea}{\begin{eqnarray}}
\newcommand{\eea}{\end{eqnarray}}
                
\newcommand{\ra}{\rightarrow}  
\newcommand{\Ra}{\Rightarrow}

\newcommand{\baa}{\begin{array}}
\newcommand{\eaa}{\end{array}}

\def\symbolfootnote[#1]#2{\begingroup
\def\thefootnote{\fnsymbol{footnote}}\footnote[#1]{#2}\endgroup}

\usepackage{hyperref}

\begin{document} 
\begin{flushright}
  $   $
\end{flushright}
\thispagestyle{empty}

\vspace{2.5cm}
\begin{center}
{\Large \bf Quantum gravity  from  Weyl conformal geometry}

 \vspace{1.5cm}
 
 {\bf D. M. Ghilencea}
 \symbolfootnote[1]{E-mail: dumitru.ghilencea@cern.ch}
 
\bigskip 

{\small Department of Theoretical Physics, National Institute of Physics
 \smallskip 

 and  Nuclear Engineering (IFIN), Bucharest, 077125 Romania}
\end{center}

\medskip

\begin{abstract}
\begin{spacing}{0.99}
  \noindent
We review recent developments in physical implications of Weyl conformal geometry.
The associated Weyl quadratic gravity action  is a  gauge theory of the Weyl group
of dilatations and Poincar\'e symmetry. Weyl conformal geometry  is defined by
equivalence classes of the metric and  Weyl gauge field ($\w_\mu$), related by
Weyl gauge transformations. Weyl geometry can be seen as  a {\it covariantised}
version of Riemannian geometry with respect to Weyl gauge symmetry (of dilatations).
This Weyl gauge-covariant formulation of Weyl  geometry  is {\it metric},
which avoids  century-old criticisms on the physical relevance of this geometry,
that ignored its gauge symmetry.
Weyl quadratic gravity and its geometry have interesting properties:
a)  Weyl gauge symmetry is spontaneously broken and Einstein-Hilbert gravity
 and Riemannian geometry are recovered,  with $\Lambda>0$;
b)~this is the only true gauge theory  of a space-time symmetry  i.e. with a
{\it physical}  (Weyl) gauge boson ($\omega_\mu$); c)  all fields and masses
have {\it geometric} origin (with no added scalar fields);
d) the theory has  a Weyl gauge invariant {\it geometric} regularisation
(by $\hat R$) in $d$ dimensions and it is Weyl-anomaly free; this anomaly is recovered
in the broken phase after massive $\omega_\mu$ decouples;
e)  the theory is the leading order of the general  Weyl gauge invariant
Dirac-Born-Infeld (WDBI) action  of Weyl conformal geometry in $d$ dimensions;
f) in the limit of vanishing Weyl gauge current, one obtains conformal gravity;
g)  finally, Standard Model (SM) has a natural embedding in conformal geometry
with no new degrees of freedom, with  successful Starobinsky-Higgs inflation.
Briefly, Weyl conformal geometry generates a  (quantum) gauge theory of gravity,
given by Weyl quadratic gravity, and leads to a unified description,
by the gauge principle, of Einstein-Hilbert gravity and SM interactions.
\end{spacing}
\end{abstract}

\noindent\begin{center}
\href{https://inspirehep.net/files/db13abdca508813b03849525464b3ec9}{\bf
Dedicated to the memory of Graham G. Ross}
\end{center}

\newpage
$   $
\vspace{4.5cm}
\tableofcontents
\newpage

\section{Motivation and brief history}\label{1}

The search for  a unified, ultraviolet (UV) complete quantum theory
of Einstein-Hilbert gravity and Standard Model (SM) largely followed
the string theory path. The   problem  with alternative approaches might not be
quantising gravity (see e.g. \cite{Donoghue}),
but finding the right underlying 4D geometry.
We thus depart from the common use of Riemannian geometry\footnote{In an abuse of language
by this  we  actually mean  pseudo-Riemannian geometry throughout this work.}
for gravity and use the Noether's gauge symmetry principle \cite{Noether} that worked so well in
SM:  the gauged {\it space-time symmetry} that we consider dictates the 4D  geometry
which dictates the  gravity action, as a gauge theory (quadratic) action. 

At  high energies and in the early Universe all physical  states are effectively massless,
which may indicate that the SM and gravity can both be part of a theory with some
form of scale symmetry\footnote{
 Nature  points  us to a self-organising principle 
with a discrete scale symmetry  known as self-similarity or self-affinity,
that exists from microscopic to astronomic scales.
This means that an affine transformation \cite{Self} that scales up 
a small region of the system  recovers a structure resembling the original,
whole system. Examples are: the brain neurons network;
snowflake structure (Koch curve) \cite{Koch};
(fractal) Brownian motion; collective systems \cite{si};  
 clustering of stars and galaxies structure formation \cite{astro}.
 These  have  self-repeating patterns at all scales with new structure
 revealed, and bring us to complex systems and fractals geometry~\cite{Mandelbrot}.
 For a fractal (non-differentiable)  geometry of the space-time  itself
 connections to fundamental physics can  emerge  \cite{N1} in an attempt to
unify large and small scale physics.
 Here we go beyond this discrete scale symmetry to study its continuous 
 version in differentiable 4D space-time geometry (Weyl conformal geometry).}:
 global, local or gauged,  where  all mass scales,
including  the Planck scale, are generated spontaneously
by vacuum expectation values of some scalar fields e.g. \cite{Misha1,t1,t2,t3,t4,Englert}.

Given the great success of gauge symmetries, to describe gravity one
usually  gauges the full conformal group \cite{Kaku}. One obtains 
{\it conformal gravity} \cite{mannheim} with an action $\sqrt{g}\,
C_{\mu\nu\rho\sigma}^2$, where $C_{\mu\nu\rho\sigma}$ is the Weyl tensor.
However, this {\it is  not} a true gauge theory of this
space-time symmetry since it has no associated  {\it physical} gauge boson
present in the spectrum,
because the action may not  have dynamical gauge bosons of special conformal 
and gauged dilatation symmetries of the conformal group \cite{Kaku}
(adding supersymmetry, leading to superconformal gravity \cite{vanProeyen},
cannot change the situation; this  raises questions how fundamental
such theory is, its link to  strings, etc). 
The underlying 4D geometry is then  Weyl {\it integrable} geometry \cite{Kaku}
conformal to Riemannian geometry, so this is not a fundamental change of  geometry.

In this work we actually seek a {\it true  gauge theory} of a space-time symmetry i.e.
with a dynamical (physical) gauge boson, in which to embed both SM and Einstein gravity.
Then the {\it only} option is to restrict  the full conformal group and study 
the gauge theory of the smaller Weyl  group (of dilatations and Poincar\'e
symmetry)\footnote{Let us also add here the remaining possibility of
gauging the Poincar\'e group only, but this brings an infinite series of higher
derivative terms and a host of complications for which we see little motivation.}
\cite{Tait}.
In such case the associated Weyl gauge boson of dilatations is indeed dynamical
and there is a (non-trivial) conserved Weyl current $J_\mu$ of 
this symmetry.  The underlying geometry, with this symmetry ``built-in'',
is then  Weyl conformal geometry
(see shortly) introduced  by Weyl a century ago \cite{Weyl1,Weyl2,Weyl3}.
The gauge symmetry principle thus demands a departure from  Riemannian
geometry-based gravity.
The associated action is the Weyl gauge theory of quadratic gravity, also known as
Weyl quadratic gravity\footnote{Misleadingly, an action given by Weyl tensor-squared is at
times  called  Weyl gravity but this is a gross  simplification; Weyl's original
theory called here Weyl quadratic gravity is so much more than that
\cite{Weyl1,Weyl2,Weyl3}.} and studied here.

Since gravity ``is'' geometry,  for an action with Weyl gauge symmetry (of dilatations)
the underlying geometry i.e. its connection is  expected to have the
(space-time) symmetry of the action. In a particular (Weyl gauge covariant) formulation of
Weyl geometry, discussed below,  the Weyl and spin  connections  have this symmetry.
This is why Weyl geometry is the right framework for our study!
The Riemannian geometry is not suitable since its Levi-Civita
connection does not have this symmetry. One could actually
insist of using the Riemannian picture or  a general affine
geometry to implement the gauged scale symmetry in an action, by using the
``standard'' tangent space formulation, uplifted to space-time by the vielbein;
however,  the formalism becomes less transparent because one is loosing
manifest Weyl gauge covariance, introduce torsion  \cite{CDA}, etc.
To understand why this happens  and for an intuitive picture,
we notice  that in  gauge theories language {\it Weyl geometry can be regarded  as a 
 covariantised version   of Riemannian geometry (the derivative of $g_{\mu\nu}$, 
connections)  with respect to the gauged dilatation symmetry.}
Moreover,   in this Weyl gauge covariant formulation  \cite{DG1},
non-metric Weyl (affine) geometry   becomes    {\it metric}
($\hat\nabla_\mu g_{\alpha\beta}\!=\!0$) and {\it non-affine!}
This is important
since then one does not need to go to a (metric) Riemannian picture, as usually done,
to do classical or quantum calculations like Weyl anomaly, etc.

In the limit the Weyl gauge  current of the theory is  trivial $J_\mu\!=\!0$,
 Weyl quadratic gravity  reduces to the earlier mentioned conformal
gravity  based on Weyl integrable geometry plus a  Weyl-invariantly
coupled  dilaton of geometric origin, that generates the  Planck scale.

A touch of history is in place here to complete the picture. While conformal
geometry is a brilliant construction of Weyl's genius,  its original physical
interpretation  as a unified  theory of  gravity {\it plus}  electromagnetism
failed - it was  later understood that electromagnetism corresponds to an  internal
symmetry of the action, not to the space-time symmetry (dilatation) of Weyl geometry.
Hence, the  Weyl gauge boson of dilatations $\w_\mu$ 
is {\it not}   the real photon as Weyl initially thought, but
a vector field that together with the metric make Weyl geometry
a vector-tensor gauge theory of gravity,  called here Weyl  quadratic gravity.

But a century ago Einstein had a different argument against any
physical relevance of Weyl geometry as  a theory of gravity \cite{Weyl2}.
He considered that the {\it non-metricity} of this geometry
by which we mean a non-zero $\tilde\nabla_\mu g_{\alpha\beta}\!\not=\!0$,
makes it unphysical. The argument was that due to this  non-metricity,
under parallel transport of a vector,  its length and in particular the
distance between the spectral lines of an atom, become
path-history dependent. This contradicts all experimental evidence
(second clock effect) \cite{Weyl2}. This view was  independent of whether
$\w_\mu$ was or  not the real photon. This argument, due to a non-zero
$\tilde\nabla_\mu g_{\alpha\beta}$ lasted a century and outlived the  attempts to
reconsider Weyl quadratic gravity  as a physical  theory of gravity.
For a review  of such attempts and early references see~\cite{Scholz}.

It is time  that this century-old  view be finally forgotten because
the original argument of Einstein simply does not apply  to a gauge theory as here.
Contrary to this  long-held view,
in Weyl geometry the length (norm) of the vector (or clock rate) 
{\it invariant} in the tangent space, remains invariant under the
parallel  transport along a curve in space-time and so there is {\it no}
path-history dependence {\it provided that  Weyl gauge-covariance is respected}
\cite{Lasenby}, \cite{non-metricity} (Appendix~B),
something actually required  in a gauge theory!
Intuitively, this result can be seen as follows:  manifest  Weyl gauge 
{\it covariance} renders Weyl geometry {\it metric} with respect to the (new)
Weyl gauge covariant derivative $\hat\nabla$, so $\hat \nabla_\mu g_{\alpha\beta}\!=\!0$,
 \cite{DG1,CDA,D2}.

Moreover,  Einstein's argument implicitly assumes a massless
Weyl gauge boson  ($\w_\mu$), which is not true: $\w_\mu$ becomes massive by a
Stueckelberg mechanism \cite{Ghilen0,SMW} and thus can decouple, and one recovers
Riemannian geometry and Einstein-Hilbert gravity, in this broken phase. Then
any non-metricity effects that one could still claim are anyway strongly
suppressed  by the mass of $\w_\mu$ which is close to Planck scale
for Weyl coupling $\alpha\sim \cO(1)$. As a side-remark, this mass
like  all masses  of the theory ($M_p$, $\Lambda$, etc),
has a {\it geometric} origin \cite{non-metricity} since it is
due to a scalar mode propagated by a (geometric) $\hat R^2$ term in the action;
no scalar fields  are added to this purpose.
Moreover, there is no need to stabilize the vev of this scalar mode 
since it is eaten by $\omega_\mu$ in a Stueckelberg mechanism.
Regarding the symmetric phase of  Weyl gauge theory of dilatations,
 a mass term is forbidden in the action. Without a mass, there is no
clock rate and  without a clock rate  there  is no second clock effect!

The main reason why   Weyl conformal geometry can be  the underlying geometry 
of a quantum gravity theory is that the gauged dilatations symmetry of Weyl quadratic
gravity is manifestly  maintained at the quantum level i.e.
it is anomaly-free \cite{DG1}. This is
required for a consistent (quantum) gauge theory of this symmetry.
Recall that in Riemannian geometry we have the famous  Weyl anomaly
\cite{Duff,Duff2,Duff3,Deser1976,Deser}. The absence of this anomaly
here is partly due to the Weyl gauge covariance in $d$ dimensions of both the
Weyl {\it and} Euler terms in the action\footnote{This enables a Weyl
gauge invariant regularisation of their contribution.} - this is not true
for the Euler term in the Riemannian case and this is where Weyl geometry differs.
Weyl anomaly is recovered in the (Stueckelberg)   broken phase of
Weyl-geometry based action after the ``dilaton'' (actually
the would-be Goldstone of gauged dilatations)
is eaten by $\w_\mu$ and decouples (together with massive $\w_\mu$),
and then Weyl geometry (Weyl connection) become Riemannian.
The absence of anomaly in a Weyl conformal  geometry-based action is
also  due to this additional dynamical degree of freedom (dof)
compared to Riemannian case.
Weyl anomaly in a  Riemannian geometry-based  gravity signals the absence
of this dof; indeed, if it is added to a conformal gravity action to enable conformal
symmetry in $d$ dimensions,  one  can obtain an anomaly-free
quantum conformal gravity \cite{Englert}.

Interestingly,  we show that Weyl quadratic gravity is 
the leading order (in some expansion) of a  more general Weyl gauge invariant
theory (in $d$ dimensions!) that we construct 
and which is a version of Dirac-Born-Infeld (DBI)
\cite{BI,D,Sorokin,Gibbons} action but in  Weyl conformal geometry.
We call this new action Weyl-DBI action (WDBI).
It is important to note that the  Weyl-DBI action  in $d$ dimensions
gives a natural regularisation (analytical continuation)
of Weyl quadratic gravity to $d$ dimensions  that is Weyl gauge
invariant and was used  in the study of Weyl anomaly. The Weyl-DBI action
also captures (apparently non-perturbative) non-polynomial
quantum corrections present in Weyl quadratic gravity.
The relation  of Weyl quadratic gravity to this Weyl-DBI
action is studied in some detail.

What happens when we add matter in the theory?
Since the SM with a vanishing Higgs mass parameter is scale invariant,
it has a  natural, minimal  embedding in Weyl geometry 
with {\it no} new  degrees of freedom needed \cite{SMW}. 
From the SM spectrum, only the  Higgs scalar ($\sigma$)
acquires tree-level couplings to $\w_\mu$, with possibly interesting phenomenology;
for example,  for a light $\w_\mu$, the annihilation  $\sigma+\sigma\ra \w_\mu+\w_\mu$ leads
to missing energy and local Weyl scalar curvature change, which is intriguing.
Successful Starobinsky-Higgs inflation is possible \cite{WI3,WI1,WI2}
being just a gauged version of
Starobinsky inflation \cite{Starobinsky}. Good fits for the galaxies
rotation curves seem possible \cite{Harko} with $\w_\mu$ as a candidate
for dark matter (implying in a sense a geometric solution to this problem);
 black hole solutions were studied in \cite{Harko2}.
Finally, the presence of the  Weyl vector in this geometry may be  needed for the
geodesic completeness \cite{Ohanian,Ehlers}, with implications for
Big-Bang and black-hole singularities.

Briefly, with its  Weyl gauge symmetry, Weyl geometry
generates a (quantum) gauge theory of gravity, given by Weyl quadratic gravity
and leads to a unified description by the gauge principle, of
Einstein-Hilbert gravity and SM interactions.

\section{Weyl geometry: the  gauge theory behind  gravity \& SM}

\subsection{Weyl geometry: a metric gauge theory with manifest  Weyl covariance}

In this section we review conformal geometry and its associated quadratic
gravity  action which is a 
gauge theory of dilatations with manifest Weyl gauge covariance, following  \cite{DG1}.

Weyl conformal geometry generalises the notion  of scale symmetry,
to a gauged version of it. More exactly, 
Weyl geometry is defined by classes of equivalence $(g_{\alpha\beta}, \w_\mu)$ of the metric
$g_{\alpha\beta}$ and the Weyl gauge boson of dilatations $\w_\mu$.
These classes are related by the  Weyl gauge  transformation shown below in
general $d=4-2\epsilon$ dimensions, in the absence (a) and the presence (b)
of scalars $\phi$ and fermions $\psi$:
\begin{equation}
  \label{WGS}
  \begin{aligned}
    (a) &\quad  g_{\mu\nu}^\prime=\Sigma^q  \,g_{\mu\nu},\qquad
    \w_\mu'=\w_\mu -\frac{1}{\alpha} \partial_\mu\ln\Sigma,  \qquad
    \sqrt{g'}=\Sigma^{q \,d/2} \sqrt{g},\quad   g^{\prime \,\mu\nu}=\Sigma^{-q}  \,g^{\mu\nu}, \\[4pt]
    (b) &\quad \phi' = \Sigma^{q_\phi} \phi, \quad
    \quad \psi'=\Sigma^{q_\psi}\,\psi,   \qquad
    q_\phi=-\frac{q}{4} (d-2);
    \qquad q_\psi=-\frac{q}{4} (d-1).
   \end{aligned}
\end{equation}

\medskip\noindent
with $\Sigma=\Sigma(x)$.
Conventions for $q$ vary: $q=1$  \cite{Smolin,Ghilen0} or $q=2$  \cite{Kugo}, etc.
If $q=2$, then for  $d=4$ one has $q_\phi=-1$,  $q_\psi=-3/2$ i.e. the inverse
mass dimensions  of $\phi$ and $\psi$.
But such normalization of an Abelian charge is a choice, hence
we keep $q$ arbitrary.  We also work in $d$ dimensions, since this is
required later for the Weyl  anomaly.

Eq.(\ref{WGS})  defines the {\it Weyl gauge symmetry} or gauged dilatations symmetry. 
Weyl geometry can thus be seen as a gauge theory of dilatations\footnote{This
  statement is more rigorously explained elsewhere using  the standard, tangent space
  formulation of the gauged dilatations  symmetry, uplifted
  to space-time by using the vielbein \cite{CDA}.}. 
Note the difference from the so-called (local) `Weyl symmetry'
where  there is no gauge field in (\ref{WGS}) ($\w_\mu\!=\!0$ or is `pure gauge').

By definition Weyl geometry is {\it non-metric} which
means $\tilde\nabla_\mu g_{\nu \rho}\not=0$, with:

\begin{equation}
  \label{tildenabla}
  (\tilde\nabla_\lambda +\alpha \,q \, \w_\lambda) g_{\mu\nu}=0, 
  \qquad \textrm{where}\qquad
  \tilde\nabla_\lambda g_{\mu\nu}
  =\partial_\lambda g_{\mu\nu}
  -\tilde\Gamma^\rho_{\lambda \mu} g_{\rho\nu}
  -\tilde \Gamma^\rho_{\lambda \nu} g_{\rho\mu}.
\end{equation}

\medskip\noindent
Since
\bea\label{fff}
\tilde\nabla_\lambda\Big\vert_{\partial_\lambda\ra \partial_\lambda+ \alpha\,q\,\w_\lambda} g_{\mu\nu}=0,
\eea
 the Weyl geometry connection $\tilde\Gamma_{\mu\nu}^\rho$,
assumed {\it symmetric} in the lower indices, is then
\medskip
\begin{equation}
  \label{tildeGamma}
  \tilde \Gamma_{\mu\nu}^\rho
  =\Gamma_{\mu\nu}^\rho\Big\vert_{\partial_\mu \ra \partial_\lambda + \alpha\,q\, \omega_\lambda}
  \!\!\!
  =\Gamma_{\mu\nu}^\rho  +\alpha^\prime \big[\delta_\mu^\rho \,\w_\nu +
    \delta_\nu^\rho \,\w_\mu-g_{\mu\nu}\, \w^\rho\big],\qquad
  \alpha^\prime\equiv\alpha\,q/2.
\end{equation}

\medskip\noindent
with $\Gamma_{\mu\nu}^\rho$ the  Levi-Civita (LC) connection of Riemannian case.
Solution (\ref{tildeGamma}) was found by a  ``covariantisation'' of the  derivative of
the metric:
$\partial_\lambda g_{\mu\nu} \ra(\partial_\lambda+\alpha\, q \omega_\lambda) g_{\mu\nu}$,
($q$= the charge of $g_{\mu\nu}$) with respect to gauged dilatations symmetry\footnote{
Note there is no complex ``i'' factor in the derivative.},
applied to $\Gamma_{\mu\nu}^\lambda$. Here
\bea
\Gamma_{\mu\nu}^\rho=(1/2)\, g^{\rho\lambda}
(\partial_\mu g_{\nu\lambda} +\partial_\nu g_{\mu\lambda}-\partial_\lambda g_{\mu\nu}).
\eea

\medskip\noindent
In other words, Weyl geometry defined by $g_{\mu\nu}$ and $\tilde\Gamma_{\mu\nu}^\lambda(g,\w_\mu)$
can be seen as a {\it covariantised  version} of Riemannian geometry (defined by
$g_{\mu\nu}$, with $\Gamma_{\mu\nu}^\lambda(g)$), with respect to the gauged dilatations symmetry.
Note that  unlike the LC connection, $\tilde\Gamma$ is invariant under (\ref{WGS}).

In what we call below  the ``non-metric formulation'' of Weyl geometry,  common
in the literature, the Riemann curvature tensor  (hereafter Weyl-Riemann tensor)
has a structure similar to that in  Riemannian geometry but in terms of
$\tilde\Gamma_{\mu\nu}^\rho$:
\bea\label{Rie}
\tilde R^\lambda_{\,\,\mu\nu\sigma}
&=&
\partial_\nu\tGamma^\lambda_{\mu\sigma}
-\partial_\sigma\tGamma_{\mu\nu}^\lambda
+\tGamma_{\nu\rho}^\lambda\,\tGamma^\rho_{\mu\sigma}
-\tGamma_{\sigma\rho}^\lambda\,\tGamma^\rho_{\mu\nu}.
\\[5pt]
&=&
R^\lambda_{\,\,\,\mu\nu\sigma}
+ \alpha^\prime
\Big\{
\delta_\sigma^\lambda \nabla_\nu \w_\mu
-\delta_\nu^\lambda\nabla_\sigma\w_\mu
-g_{\mu\sigma}\nabla_\nu \omega^\lambda
+ g_{\mu\nu} \nabla_\sigma\omega^\lambda
+\delta_\mu^\lambda \,F_{\nu\sigma}\Big\}\nonumber
\\[3pt]
&+&  \alpha^{\prime 2}
\Big\{
\omega^2 (\delta_\sigma^\lambda\,g_{\mu\nu} 
-\delta_\nu^\lambda \,g_{\mu\sigma})
+\w^\lambda \,(\w_\nu g_{\sigma\mu}-\w_\sigma g_{\mu\nu})
+ \w_\mu \,(\w_\sigma \delta_\nu^\lambda -\w_\nu\,\delta_\sigma^\lambda)\Big\}\label{A6}
\label{RIE}\eea
where $R^\lambda_{\,\,\,\mu\nu\sigma}$ is the Riemannian counterpart and
$F_{\nu\sigma}=\tilde\nabla_\nu\w_\sigma-\tilde\nabla_\sigma\w_\nu
=\partial_\nu \w_\sigma-\partial_\sigma\w_\nu$.
In (\ref{RIE}) we used relation (\ref{tildeGamma}).
Note in the rhs
the term $\delta_\mu^\lambda F_{\nu\sigma}$ which seems out of place (see later).
From (\ref{RIE}) we derive the  Ricci tensor  $\tilde R_{\mu\nu}\equiv\tilde R^\rho_{\,\,\mu\rho\nu}$
(Weyl-Ricci), Weyl scalar $\tilde R$  and Weyl tensor $\tilde C_{\mu\nu\rho\sigma}$ 
of Weyl geometry
\begin{align}
  &\tilde R_{\mu\nu}= 
   R_{\mu\nu} +
   \alpha^\prime \Big[
   \frac{d}{2} F_{\mu\nu}-(d-2)\nabla_{(\mu} \w_{\nu)}
   - g_{\mu\nu} \nabla_\lambda\w^\lambda\Big]
   +
   \alpha^{\prime \, 2}
   (d-2) (\w_\mu\w_\nu -g_{\mu\nu} \w_\lambda\w^\lambda), \label{Ri}
  \\[6pt]
  &\,\,\,\,\,\, \tilde R = g^{\mu\nu}\tilde R_{\mu\nu}= R-2 (d-1)\,\alpha^\prime\,
  \nabla_\mu \w^\mu  -(d-1) (d-2)\,\alpha^{\prime\,2}\, \w_\mu \w^\mu,\label{Ri2}
\\
&\!\!\! \tilde C_{\mu\nu\rho\sigma}\!=\!
\tilde R_{\mu\nu\rho\sigma}
+\frac{1}{d-2} \big( g_{\mu\sigma} \tilde R_{\nu\rho}+g_{\nu\rho} \tilde R_{\mu\sigma}
-g_{\mu\rho} \tilde R_{\nu\sigma}-g_{\nu\sigma} \tilde R_{\mu\rho}\big)
+\frac{ \big(g_{\mu\rho}g_{\nu\sigma}-g_{\mu\sigma} g_{\nu\rho}\big)\tilde R}{(d-1)(d-2)}.
\end{align}

\medskip\noindent
Here  $ R_{\mu\nu}$, $ R$ denote the Ricci tensor and scalar in the Riemannian case;
$\nabla$ is  the Riemannian geometry derivative 
$\nabla_\mu \omega_\nu=(\partial_\mu\omega_\nu-\Gamma_{\mu\nu}^\rho \omega_\rho)$ 
and  $\nabla_{(\mu}  \w_{\nu)}\equiv (1/2)\,(\nabla_\mu\w_\nu+\nabla_\nu\w_\mu)$.
Regarding the Weyl tensor $\tilde C_{\mu\nu\rho\sigma}$,
it has the property   $\tilde C_{\mu\nu\rho\sigma}^2=C_{\mu\nu\rho\sigma}^2+
3\,\kappa\, F_{\mu\nu}^2$ where $\kappa$ is a numerical coefficient, see eq.(A-14) in \cite{DG1},
with $C_{\mu\nu\rho\sigma}$ the Weyl tensor of Riemannian geometry.

Given the invariance of $\tilde \Gamma$ under (\ref{WGS}),
the Weyl-Riemann and Weyl-Ricci 
tensors are also invariant while the Weyl scalar $\hat R$ 
is covariant (due to $g^{\mu\nu}$ in its formula):
\medskip
\bea\label{tra}
\tilde R^{\prime\mu}_{\,\,\,\,\nu\rho\sigma}=\tilde R^\mu_{\,\,\,\,\nu\rho\sigma},
\quad
\tilde R^{\prime}_{\mu\nu}=\tilde R_{\mu\nu},
\quad
\tilde R^\prime=\Sigma^{-q} \tilde R,
\quad
\tilde C_{\mu\nu\rho\sigma}'=\Sigma^q\,\tilde C_{\mu\nu\rho\sigma},
\quad
\tilde F_{\mu\nu}^\prime=\tilde F_{\mu\nu}.
\eea

\medskip\noindent
In a Riemannian case, some of these terms
transform under (\ref{WGS}) in a very complicated way.

With the above formulae  one can write the most general
action in Weyl geometry that is Weyl gauge invariant. This action contains
the following independent terms written by Weyl more than a century ago  \cite{Weyl2}
in an equivalent basis of operators
\medskip
\bea\label{S}
S_{\bf }=\int d^4x  \sqrt{g}\,\Big\{ a_0\tilde R^2+ b_0 \tilde F_{\mu\nu}^2+
c_0 \tilde C_{\mu\nu\rho\sigma}^2 +d_0\, \tilde G\Big\},
\eea

\noindent
with some dimensionless coefficients $a_0, b_0, c_0, d_0$.
$\tilde G$ is the Chern-Euler-Gauss-Bonnet term (hereafter Euler term)
of Weyl geometry which
is a total derivative in $d=4$ but relevant
in $d$ dimensions; the expression of $\tilde G$ is found in eq.(A-15)
of \cite{DG1}\footnote{There exists another topological term,  the Pontryagin term which
is parity-odd (so it is usually ignored) but unlike $\hat G$,
it does not exist in $d$ dimensions
\cite{CDA2} (section 2.5); in fact there are two such terms  \cite{Frolov}.}.
Each term in the action is separately Weyl gauge invariant, as one can easily check.

This is the most general action that can be used for a gravity theory based on Weyl geometry.
The are  no higher dimensional terms here,  since there is no mass scale
in the action to suppress them. One can also consider adding (invariant)
non-polynomial terms such as
$\sqrt{g}\,(\tilde C_{\mu\nu\rho\sigma}^2)^2/\tilde R^2$, etc. We discuss them later on.
One could  then analyse action (\ref{S}), but since geometry seems non-metric, one must
go to a Riemannian picture to do this.

\bigskip
\noindent{\bf $\bullet$ Weyl gauge covariant formulation is metric}

\medskip
However, as already remarked,  Weyl geometry has a gauged 
dilatations symmetry (\ref{WGS}), hence we would like a formulation of this geometry with manifest
Weyl gauge-covariance, as in any gauge theory. But a
derivative like $\tilde \nabla_\mu \tilde R$ is not Weyl gauge-covariant under (\ref{WGS}),
since $\tilde \nabla_\mu^\prime \tilde R^\prime\not=\Sigma^{-q}\tilde \nabla_\mu \tilde R$,
and similar for other fields.
So we must upgrade to a different,   Weyl gauge-covariant ``basis'' and differential operator.
To do so, notice that $(\tilde\nabla_\lambda +\alpha\, q \, \w_\lambda) g_{\mu\nu}=0$
where $q$ is the
space-time charge of the metric;
this suggests that for any given tensor $T$ of {\it space-time charge}
$q_T$,  in particular $g_{\mu\nu}$,  with $T^\prime=\Sigma^{q_T} T$,  one should define
a new differential operator $\hat \nabla_\mu$ that replaces
$\tilde\nabla_\mu$:
\medskip
\bea\label{qq}
\hat \nabla_\mu T
=\tilde\nabla_\mu\Big\vert_{\partial_\mu\ra \partial_\mu + \alpha\, q_T\, \w_\mu} T
\equiv (\tilde\nabla_\mu +\alpha\, q_T \, \w_\mu) T\qquad
\Ra\qquad \hat\nabla_\mu' T'=\Sigma^{q_T}\, \hat\nabla_\mu T.
\eea

Hence,  $\hat\nabla_\mu$ transforms covariantly under (\ref{WGS}),
as seen by using
that $\tilde\Gamma$ is invariant. Eq.(\ref{qq}) simply introduces a Weyl gauge covariant
operator
$\hat\nabla_\mu$  by a  ``covariantisation'' of the partial derivative in $\tilde\nabla_\mu$:
$\partial_\mu\ra\partial_\mu + \alpha\times \rm{charge} \times\w_\mu$.
Note that  no $\hat\Gamma$ can be associated to $\hat\nabla$ because the charge $q_T$
depends on the field $T$ that $\hat\nabla$ acts on, hence this is {\it not an affine} formulation.
This new formulation is however {\it metric} (with respect to the new operator $\hat \nabla_\mu$), since
\bea\hat\nabla_\mu g_{\alpha\beta}=0.
\eea

One can then proceed and define new  Weyl-Riemann and Weyl-Ricci tensors by
using the new operator $\hat\nabla_\mu$ in the commutator that defines
the Weyl-Riemann tensor \cite{CDA}, as in 
\bea\label{rrr}
    [\hat\nabla_\mu,\hat\nabla_\nu] v^\rho=\hat R^{\rho}_{\,\,\,\sigma\mu\nu} v^\sigma
    \eea
where  $v^\rho$ is a vector of vanishing Weyl charge on the tangent space \cite{CDA}.
Physically, using invariant vectors on the  tangent space to define $\hat R^\rho_{\,\,\,\sigma\mu\nu}$
is  the most natural choice.
The same result for the new curvature tensors (with a hat) is also  obtained by simply
re-defining  $\tilde R_{\mu\nu\rho\sigma}$ and $\tilde R_{\nu\sigma}$ as below
\cite{DG1,Jia}
\be\label{nat}
\hat R_{\mu\nu\rho\sigma}=\tilde R_{\mu\nu\rho\sigma}-\alpha^\prime \, g_{\mu\nu} \hat F_{\rho\sigma},
\quad
\hat R_{\nu\sigma}=\tilde R_{\nu\sigma}-  \alpha^\prime\,\hat F_{\nu\sigma},
\quad
\hat R=\tilde R.
\ee

\medskip\noindent
The term subtracted above
is the ``odd'' term $\delta_\mu^\lambda F_{\nu\sigma}$  in (\ref{RIE}).
Here  $\hat F_{\mu\nu}\!=\!\hat\nabla_\mu \w_\nu-\hat\nabla_\nu\w_\mu=
\partial_\mu \w_\nu-\partial_\nu \w_\mu\!=F_{\mu\nu}$. Note also
that $\hat R_{\mu\nu}-\hat R_{\nu\mu}\!=\alpha^\prime\,(d-2) F_{\mu\nu}$, used later.
With (\ref{Rie}), (\ref{Ri})  one immediately writes the tensor/scalar curvatures
in hat basis in terms of their Riemannian version. 
 From (\ref{nat}) the  Weyl tensor  due to $\hat R_{\mu\nu\rho\sigma}$   and the Euler
 term become \cite{DG1} (Section~3.1) in $d$ dimensions:
 \bea\label{ccc}
\hat C_{\mu\nu\rho\sigma}= C_{\mu\nu\rho\sigma},
\qquad
\hat G= \hat R_{\mu\nu\rho\sigma} \hat R^{\rho\sigma\mu\nu}
- 4 \hat R_{\mu\nu} \hat R^{\nu\mu} +\hat R^2.
\eea

\medskip\noindent
Hence the Weyl tensor $\hat C^\lambda_{\,\,\,\mu\nu\rho}$
is identical to that in Riemannian case, $C^\lambda_{\,\,\,\mu\nu\rho}$, while
$\hat G$ is the Euler term $\tilde G$ written in this covariant ``hat basis''.
Note the position of summation indices.

Under (\ref{WGS}) we have the same transformations
in the ``hat basis''/formulation 
as in (\ref{tra}):
\medskip
\bea\label{tra2}
\hat R^{\prime\mu}_{\,\,\,\,\nu\rho\sigma}=\hat R^\mu_{\,\,\,\,\nu\rho\sigma},
\qquad
\hat R^{\prime}_{\mu\nu}=\hat R_{\mu\nu},
\qquad
\hat R^\prime=\Sigma^{-q} \hat R,
\qquad
\hat F_{\mu\nu}^\prime=\hat F_{\mu\nu}.
\eea

However, under (\ref{WGS})
the ``hat'' formulation is {\it Weyl gauge covariant}  in
$d$ dimensions
\medskip
\be\label{WGS3}
 X^\prime=\Sigma^{-2\, q} X, \qquad
 X =\hat R_{\mu\nu\rho\sigma}^2, \, \,\hat R_{\mu\nu}^2,\,\,\hat R^2,
 \,\, \hat C_{\mu\nu\rho\sigma}^2,\,\, \hat G,
\,\, \hat F_{\mu\nu}^2.
\ee
\be\label{wq}
\hat\nabla_\mu' \hat R^\prime=\Sigma^{-q}\,\hat \nabla_\mu \hat R,\qquad 
\hat\nabla^\prime_\mu\hat\nabla^{\prime}_\nu \hat R'
=\Sigma^{-q}\, \hat \nabla_\mu\hat\nabla_\nu \hat R, \quad 
\hat\nabla_\rho' \hat R_{\mu\nu}'=\hat\nabla_\rho \hat R_{\mu\nu},\,\,\,
\text{etc.}
\ee

\medskip
Hence both the curvature tensors/scalar {\it and} their derivatives $\hat\nabla_\mu$  
transform covariantly, unlike in the non-metric formulation of $\tilde\nabla_\mu(\tilde\Gamma)$.
We now have a  {\it metric, Weyl gauge covariant formulation} (with respect to $\hat\nabla_\mu$)
 that replaced the non-metric formulation of $\tilde\nabla_\mu(\tilde\Gamma)$.
Note that no $\hat\Gamma$ can be associated to $\hat\nabla_\mu$ in this case due to the different
 charges of operators, hence the Weyl gauge covariant formulation is not an affine formulation.
 
The covariant transformations (\ref{WGS3}), (\ref{wq})
are in $d$ dimensions, which is  important
 at the quantum level. In particular, the Euler term $\hat G$  is 
 Weyl gauge covariant in arbitrary $d$ dimensions, just like  $\hat C_{\mu\nu\rho\sigma}^2$.
 This is relevant  when we discuss  Weyl anomaly in conformal geometry, which requires
 a regularisation in $d$ dimensions and where  Weyl gauge covariance is used;
 this property of $\hat G$  is specific to Weyl conformal geometry.
 The situation is different from Riemannian geometry where the Euler term in
 $d$ dimensions transforms in a complicated way, with implications
 for  Weyl anomaly.
 
 For later reference, we show two relations in the ``hat'' formulation. We have \cite{DG1}
\medskip
\bea\label{id1}
\hat C_{\mu\nu\rho\sigma}^2= \hat R_{\mu\nu\rho\sigma} \hat R^{\rho\sigma\mu\nu}
-\frac{4}{d-2} \hat R_{\mu\nu} \hat R^{\nu\mu} +\frac{2}{(d-1)(d-2)}\hat R^2.
\eea

\medskip\noindent
With second equation in  (\ref{ccc}) we eliminate the Riemann term
from (\ref{id1}), to find
\medskip
\bea\label{id2}
\hat R_{\mu\nu} \hat R^{\nu\mu}=\frac{d-2}{4\, (d-3)} \big( \hat C_{\mu\nu\rho\sigma}^2 -\hat G\big)
+\frac{d}{4\, (d-1)} \hat R^2.
\eea

\medskip\noindent
These two relations extend to Weyl geometry identical relations in Riemannian geometry.

 In this new, Weyl gauge covariant formulation (with a ``hat'') the most general action  in
 Weyl conformal geometry is quadratic in curvatures (as in a gauge theory)
 and is given by
\medskip
\bea\label{Sf0}
S_{\bf w}=\int d^4x  \sqrt{g}\,\Big\{ a_0\hat R^2+ b_0 \hat F_{\mu\nu}^2+
c_0 \hat C_{\mu\nu\rho\sigma}^2 +d_0\, \hat G\Big\}.
\eea

\medskip
Each term in $S_{\bf w}$  is separately invariant under (\ref{WGS}) for $d=4$.
Action $S_{\bf w}$ is identical to that in the  non-metric formulation eq.(\ref{S}),
up to a redefinition of $b_0$. This is seen as follows:
$\hat R=\tilde R$, $\hat F_{\mu\nu}=\tilde F_{\mu\nu}$ and $\hat G$ and $\tilde G$ are
total derivatives. The only difference is that $\tilde C_{\mu\nu\rho\sigma}^2=C_{\mu\nu\rho\sigma}^2+
3\kappa F_{\mu\nu}^2$ where $\kappa$ is a numerical coefficient \cite{DG1}. Hence,
a redefinition of  $b_0$ shows that the two actions and formulations are equivalent.
One can then  work  either in  the non-metric formulation (of $\tilde\nabla$)
or in the metric Weyl gauge-covariant one (of $\hat \nabla$) \cite{CDA}.

Action (\ref{Sf0}) is the most general one can write in Weyl geometry because any
higher dimensional terms are forbidden since there is
no mass scale in the theory to suppress them; (or they are of
non-polynomial/non-perturbative nature, see discussion after eq.(\ref{S})).

In action (\ref{Sf0}), the Weyl gauge covariance of each term 
enables us to {\it maintain the  Weyl gauge symmetry in $d=4-2\epsilon$ dimensions},
by a natural {\it geometric} analytic continuation 
\bea\label{Sf}
S_{\bf w}=\int d^d x  \sqrt{g}\,\Big\{ a_0\,\hat R^2+ b_0 \,\hat F_{\mu\nu}^2+
c_0 \,\hat C_{\mu\nu\rho\sigma}^2 +d_0 \,\hat G\Big\} \, (\hat R^2)^{(d-4)/4}.
\eea
%
This regularisation  implicitly assumes $\hat R\not=0$, which is 
verified a-posteriori \cite{DG1}, see later. There is {\it no need for an artificial
UV regulator/subtraction scale $\mu$} as in the DR scheme - that
would break the Weyl gauge symmetry.
Weyl geometry itself, via its scalar curvature ($\hat R^\epsilon$),
acts as UV regulator and ensures Weyl gauge invariance of $S_{\bf w}$ in $d$ dimensions!
 Quantum calculations can then be done using (\ref{Sf}),
in this  Weyl gauge covariant {\it metric} formulation.

To summarize:  the ``hat''  formulation for Weyl geometry is very important:
it  enables manifest Weyl gauge covariance for
all field operators and  their derivatives - this was  obtained by a 
covariantisation of  $\tilde\nabla$ when acting on such
fields (leading to $\hat\nabla$).
In particular,  the Weyl connection eq.(\ref{tildeGamma})
is also a ``covariantised'' version of Levi-Civita connection of Riemannian geometry,
with respect to the gauged scale symmetry.
With respect  to the new  Weyl gauge covariant $\hat\nabla$,
we now have  a  (non-affine!) metric and physical formulation (replacing the initial affine,
but not Weyl-covariant and non-metric formulation of $\tilde\nabla$).
This enables us to do all calculations in Weyl geometry \cite{DG1}, which is important.
Thus, we do not have to express the Weyl action in the (metric) Riemannian geometry
notation, as required when using the non-metric formulation of Weyl geometry.

Briefly, we have a manifestly Weyl gauge-covariant metric
formulation of Weyl geometry  that has an associated quadratic gravity action
which is a gauge invariant theory (of the Weyl group)  valid in arbitrary
$d$ dimensions, eq.(\ref{Sf}).

\subsection{Tangent space formulation of Weyl gauge symmetry has torsion}

In addition to the above two formulations of Weyl geometry,
there exists a third equivalent formulation. Let us detail. For a better understanding,
in \cite{CDA} (section 2.3) we used the standard  tangent space-time
approach to construct  a  gauge theory of the Weyl group (of dilatations
and Poincar\'e symmetry), uplifted to curved space-time by the vielbein, see e.g. \cite{vanProeyen}.
The results  show that the corresponding gauge connection $\Gamma_{\mu\nu}^{' \rho}$ obtained in such
approach  relates to $\tilde\Gamma_{\mu\nu}^\rho$ of (\ref{tildeGamma}) by a projective transformation:
\bea
\tilde\Gamma_{\mu\nu}^\rho=\Gamma_{\mu\nu}^{' \rho}+\alpha^\prime\,\delta_\mu^\rho\,\omega_\nu.
\eea
%
As a result, one can check that
this formulation is metric but the initial  vectorial non-metricity (of $\tilde\nabla$)
is replaced by vectorial torsion given by $\Gamma_{\mu\nu}^{' \rho}-\Gamma_{\nu\mu}^{'\rho}\not=0$,
since unlike $\tilde\Gamma_{\mu\nu}^\rho$,  $\Gamma_{\mu\nu}^{'\rho}$ is not symmetric in the lower
indices.
The corresponding Riemann tensor ($R^{'\lambda}_{\,\,\mu\nu\sigma}$) defined as in (\ref{Rie}) but
with $\tilde\Gamma_{\mu\nu}^\rho$
replaced by $\Gamma_{\mu\nu}^{'\rho}$ is
\medskip
\bea\label{Rie2}
R^{'\lambda}_{\,\,\mu\nu\sigma}
&=&
\partial_\nu\Gamma^{'\lambda}_{\mu\sigma}
-\partial_\sigma\Gamma_{\mu\nu}^{'\lambda}
+\Gamma_{\nu\rho}^{'\lambda}\,\Gamma^{'\rho}_{\mu\sigma}
-\Gamma_{\sigma\rho}^{'\lambda}\,\Gamma^{'\rho}_{\mu\nu}.
\eea

\medskip\noindent
One checks that this tensor is actually identical to that
in the Weyl-covariant (``hat'') formulation eqs.(\ref{rrr}), (\ref{nat}),
so
\bea
R^{'\lambda}_{\,\,\,\,\mu\nu\sigma}={\hat R}^\lambda_{\,\,\mu\nu\sigma}
\eea
As a result, the  action it leads to is identical to (\ref{Sf0}).
This gives a third, equivalent formulation (metric, but with vectorial torsion)
of Weyl  geometry, but manifest Weyl gauge covariance of the new derivatives (defined  by
$\Gamma_{\mu\nu}^{'\rho}$) is now lost.
This equivalence also  shows that in Weyl geometry  one
cannot distinguish between vectorial non-metricity and vectorial
torsion\footnote{
The equivalence vectorial torsion vs vectorial non-metricity formulation
fails in general, beyond Weyl geometry, since non-metricity and torsion tensors
have different  degrees of freedom and
physical meaning; only their vector component under $so(4)$
decomposition can be seen as either torsion or non-metricity \cite{CDA}}
and all three formulations are equally good. The equivalent formulations
correspond to different combinations of generators of the Weyl group
algebra. Unlike the manifestly Weyl gauge covariant (metric) formulation,
the other two do not guarantee physical results
when gauge symmetry is relevant, as here. Since we are interested in the
Weyl-covariant and metric formulation, we do not detail further this third formulation,
for details see \cite{CDA,CDA2}.

\subsection{Weyl gauge symmetry breaking: from Weyl to Einstein-Hilbert}
\label{wgsb}

The  Weyl quadratic action in (\ref{Sf0}) and (\ref{Sf}) is  an Abelian gauge theory of dilatations
 in the absence of matter. In this section  we review  how Einstein-Hilbert  gravity is
recovered from this action after a  spontaneous  breaking of its Weyl gauge symmetry,
by following \cite{Ghilen0}, \cite{SMW}.

Consider then  action (\ref{Sf0}) in $d\!=\!4$ with  perturbative values for its
couplings, as below:
\medskip
\bea\label{inA}
S_{\bf w}=\int d^4x \sg\, \,\Big[\, \frac{1}{4!}\,\frac{1}{\xi^2}\,\hat R^2  - \frac{1}{4}
\, \hat F_{\mu\nu}^{\,2}  - \frac{1}{\eta^2}\,\hat C_{\mu\nu\rho\sigma}^{\,2}
 +d_0\,\hat G
\Big],
\eea

\medskip\noindent
with  perturbative couplings $\xi,\, \eta, \,\alpha\leq 1$.
We ignore for now the last term $\hat G$  which is topological (total derivative
in $d=4$) but becomes relevant  when we discuss Weyl anomaly.

In $S_{\bf w}$ replace  $\hat R^2\ra -2\phi^2 \hat R-\phi^4$ with $\phi$ a scalar field.
This  step gives a new $S_{\bf w}$ classically equivalent to (\ref{inA}):
indeed, by using the solution $\phi^2=-\hat R$, ($\hat R<0$)
of the equation of motion of $\phi$ in the new  $S_{\bf w}$, one recovers
initial action (\ref{inA}).
Then using (\ref{Ri2}), (\ref{nat}), (\ref{ccc}) for $d=4$  and  some re-arrangement,
the action becomes in the {\it Riemannian notation} (with $R$, $\nabla$) \cite{Ghilen0,SMW}:
\smallskip
\begin{multline}\label{alt2}
\!\! S_{\bf w}\!=\!\!\int\!\! d^4x \sqrt g
\Big\{\frac{-1}{2\xi^2} \Big[ \frac16 \phi^2\,R 
  +(\partial_\mu\phi)^2  - \frac{\alpha\, q}{2}
  \nabla_\mu (\w^\mu\phi^2)\Big]
-\frac{\phi^4}{4!\,\xi^2} + \frac{\alpha^2 q^2}{8\,\xi^2}\,\phi^2 \Big[\w_\mu
-\frac{1}{\alpha q}\partial_\mu \ln\phi^2\Big]^2\qquad
\\[3pt]
 \quad
-\frac{1}{4}\,F_{\mu\nu}^2\,-\,\frac{1}{\eta^2}\,C_{\mu\nu\rho\sigma}^2
\Big\},
 \end{multline}

\medskip\noindent
Every bracket with coefficient $\propto 1/\xi^2$ and each of the remaining terms (multiplied by
$\sqrt g$) is invariant under  (\ref{WGS}).
Next, let us apply a specific form of transformation (\ref{WGS}) that is scale-dependent
$\Sigma=\phi^2/\langle\phi^2\rangle$; this is simply fixing $\phi$ to its vev,
assumed to exist and generated e.g. at quantum level.
Naively, one simply sets $\phi\ra \langle\phi\rangle$ in the above action. In
terms of the transformed (``primed'')  fields, we obtain the broken phase action 
\medskip
\bea
\label{EP}
S_{\bf w}=\int d^4x
\sqrt{g'}  \,\Big[- \frac12\, M_p^2 \, R'
 +\frac12 m_\omega^2 \w_\mu'  \w^{\prime \mu}
 - \Lambda\, M_p^2
 -\frac{1}{4} \, F_{\mu\nu}^{' 2}-\frac{1}{\eta^2}\,  C_{\mu\nu\rho\sigma}^2
 \Big],
\eea
with the notation
\bea\label{la}
\Lambda\equiv \frac14\,\langle\phi\rangle^2,\qquad
M_p^2\equiv \frac{\langle\phi^2\rangle}{6\,\xi^2},\qquad
m_\omega^2\equiv \frac32 \alpha^2\,q^2\,M_p^2,
\eea
%
and where we identified  $M_p$ with the Planck scale and $\Lambda$ with the cosmological constant.
Note the mass of $\w_\mu$ depends on the Weyl charge normalization of the metric!

Eq.(\ref{EP}) is the Einstein gauge (``frame'') and
the {\it unitary, physical gauge} of the Weyl gauge invariant action (\ref{alt2}).
By  a  Stueckelberg  mechanism  \cite{ST}, $\w_\mu$ has become a massive Proca field, 
by ``eating'' in (\ref{EP}) the derivative $\partial_\mu\ln\phi$ of the Stueckelberg field $\ln\phi$
\cite{Ghilen0}.
Indeed, $\ln\phi$ transforms with a shift under (\ref{WGS}) hence it plays the role
of the would-be-Goldstone of this symmetry (or ``dilaton'').
Note  that the number of degrees of freedom (dof) is conserved (as it should in
spontaneous breaking),
since transformation (\ref{WGS}) cannot change the dof number: indeed,
in addition to the graviton,  the real  massless $\phi$ (dof=1) and
massless $\w_\mu$ (dof=2) were replaced by a massive Weyl-Proca field $\w_\mu$ (dof=3)
of mass $m_\w$  in eq.(\ref{EP});
($\phi$ is indeed  a dynamical field, as seen from the equations of motion,
 eq.(\ref{jj}) and Appendix B in \cite{SMW}).

The Einstein-Proca action in (\ref{EP}) is  a  broken phase of Weyl's original gauge theory of
gravity (\ref{inA}). With  $\w_\mu$ massive, it decouples from  (\ref{EP}) and below $m_\w$  
the Einstein-Hilbert action  is found as a `low-energy' effective  theory of
Weyl's theory \cite{Ghilen0}. Hence, Einstein gravity appears to
be the  ``Einstein gauge''-fixed version of the Weyl  action.  However,
 the  breaking is much  more profound and is {\it not}  the  result of
a simple  `gauge choice':  it is  accompanied by
a  Stueckelberg mechanism and  by a  transition from  Weyl  to Riemannian geometry:
indeed, when massive $\w_\mu$ decouples then Weyl connection $\tilde\Gamma$ 
is replaced by Levi-Civita connection:
\bea\label{omega-limit}
\w_\mu\ra 0,\,\qquad\qquad \tilde\Gamma_{\mu\nu}^\lambda\ra \Gamma_{\mu\nu}^\lambda,
\eea
and hence Weyl geometry becomes Riemannian.
Therefore, the breaking of Weyl gauge symmetry and subsequent  decoupling of massive $\w_\mu$ has
a {\it geometric interpretation:} it is a  transition from Weyl conformal
geometry to Riemannian geometry!

The  mass scales generation that
accompanies this spontaneous breaking  is a geometric effect:
the  cosmological constant ($\Lambda$),
Planck scale  $M_p$,  $m_\w$, are all generated by the vev of
same scalar field  $\phi$, which has a geometric origin in the $\hat R^2$ term.
The same is true about the mass $m_\eta=\eta M_p$ of the massive spin-two state
that is generated  by the  term $C_{\mu\nu\rho\sigma}^2$ in the presence of
the Einstein-Hilbert term, in action (\ref{EP}),
for details see \cite{LAG}. For a natural value  $\eta\sim \cO(1)$
that we assume in this work, then $m_\eta\sim M_p$ and it decouples below this scale. 
Briefly, all  mass scales of the theory $\Lambda$, $m_\w$, $M_p$, $m_\eta$
have  a geometric origin - no matter fields, Weyl scalar field compensators, etc,
 were  considered or added to this purpose!

The mass $m_\w$ of $\w_\mu$ depends on the Weyl gauge symmetry coupling $\alpha$,
which is not constrained. In principle  $\alpha\sim \cO(1)$,
in which case  $m_\w$  is close to the Planck scale, $m_\w\sim M_p$.
Then the Weyl boson decouples near  this scale. But one can have
$\q\! \ll\! 1$, in which case $m_\w\!\ll\! M_p$, for example $m_\w\sim$ few TeV!
Currently the only lower experimental bound on $m_\w$ is $\sim 1$ TeV found in 
\cite{Latorre} from  a process of Bhabha scattering mediated by $\w_\mu$.
The problem with this bound is that it  assumes  an interaction of fermions with $\w_\mu$
which  does not exist in our case of Weyl conformal geometry:  while fermions are charged under
the Weyl gauge group, see (\ref{WGS}), they have  no tree-level coupling to $\w_\mu$!
(see later and  \cite{Kugo,SMW}). Hence, this bound can be even lower than 1 TeV,
which is interesting for phenomenology and should be explored further. Therefore, 
an ultraweak Weyl gauge coupling $\q\ll 1$ is possible.

Further, since the Planck scale  $M_p$ is related to  $\Lambda$, the latter 
cannot vanish - in fact $\Lambda >0$. With   $M_p$ fixed, the cosmological constant
$\Lambda$ is then small because gravity is ultraweak ($\xi\ll 1$) as seen from
\bea
\frac{\Lambda}{M_p^2}=\frac32 \,\xi^2.
\eea
This suggests a possible connection UV versus IR  physics, represented by
$M_p$ and $\Lambda$ scales, respectively. 
For $\langle\phi\rangle\ra 0$ then $\Lambda, M_p\ra 0$ and  Weyl
gauge symmetry is restored and $\Lambda$ is protected by this symmetry.
This limit is rather formal, since we linearised the quadratic action in (\ref{inA})
with $\phi^2=-\hat R$ which implicitly  assumes that $\phi$ is non-zero.
Note also that for a Friedmann-Lema\^itre-Robertson-Walker
metric, one can show that 
$\hat R=-12 H_0^2<0$ and $\Lambda=3 H_0^2$, where $H_0$ is the Hubble constant
\cite{SMW,SMW2}.

To conclude,  Weyl quadratic gravity eqs.(\ref{Sf0}), (\ref{Sf}), (\ref{inA})
is more fundamental than the Einstein-Proca action (\ref{EP}) which is only  its 
``low-energy'' effective theory limit, a  broken phase in the {\it unitary gauge}
of the gauged dilatation symmetry (this is similar to the
unitary gauge of $U(1)_Y$ of SM).
When the massive Weyl gauge boson decouples, the  geometry becomes Riemannian and
 Einstein gravity is recovered. The scale at which $\w_\mu$ decouples is not
 fixed however by the theory - it depends on the Weyl gauge coupling $\alpha$ which, as
 a gravitational coupling, can be ultraweak.
This picture is entirely {\it geometrical} \cite{non-metricity}  since we did
not yet include  matter fields, and all mass scales above have  a geometric
origin.  These results suggest that the underlying geometry of our
Universe is actually  the more general  Weyl conformal geometry
rather than Riemannian geometry.

\subsection{Non-metricity? what non-metricity?}

If Einstein-Hilbert gravity is the low energy
broken phase of Weyl gauge theory of quadratic gravity,
what about the long-held criticism of Einstein that  the non-metricity of the underlying
Weyl geometry makes this theory  unphysical \cite{Weyl2}?
Let us then  review the effect of non-metricity in Weyl geometry to show that this
long-held criticism  is avoided.

By definition, non-metricity of Weyl geometry means that
\bea
\tilde\nabla_\mu g_{\alpha\beta}\not =0.
\eea
where $\tilde\nabla$ is the derivative operator of Weyl geometry (determined by $\tilde\Gamma$),
see eq.(\ref{tildenabla}).

The traditional, long-held view is that due to this relation,
in Weyl geometry, under parallel transport,
vectors change not only direction (as in the Riemannian case) but also their norm; therefore,
their norm and clock rates, too,  depend on the path history of these vectors (second-clock effect).
In particular, the distance  between the atomic spectral lines of some parallel transported atoms
would be  path-dependent, in contrast with the experiment. Thus  theories based on Weyl
geometry would be unphysical. This is the original argument of Einstein \cite{Weyl2} and
applies regardless of the fact that $\w_\mu$ is not the real photon (as Weyl had initially thought).
This long-held view is missing two key aspects:

{\bf a)} firstly, the norm of a vector $u^\mu$ (invariant in the tangent space)
does not change under parallel transport in Weyl geometry if this transport
is correctly realised i.e. preserves the Weyl gauge covariance,
as it  should in a gauge theory. Indeed, 
in the gauge covariant formulation, Weyl geometry is metric $\hat\nabla_\mu g_{\alpha\beta}=0$,
so the norm is obviously invariant.

{\bf b)}  secondly, the second clock effect demands the presence
of a {\it mass scale} in the action, but there is no mass scale in the
symmetric phase of the theory, while in the broken phase the effect is suppressed
by $m_\w$ \cite{Ghilen0}. Let us detail a) and b).

Regarding {\bf a)} above, as discussed elsewhere \cite{non-metricity} (Appendix B) and \cite{Lasenby},
the norm of a vector that is invariant  in the tangent space (in order to be physical) 
i.e. has a vanishing Weyl charge, does remain invariant under parallel transport
along a curve $\gamma(\tau)$. To see this, consider a vector
$u^\mu$ with  Weyl charge $z_u/2$, hence it transforms as
\bea
u^{\prime \mu} = \Sigma^{z_u/2} \,u^\mu.
\eea
Under parallel transport that preserves  Weyl covariance, as demanded in a gauge
theory in order to be physical, we have, following Appendix B in \cite{non-metricity} 
\bea\label{gg}
\frac{D u^\mu}{d \tau}=0, \qquad
D u^\mu\equiv
 dx^\lambda \tilde\nabla_\lambda u^\mu
\Big\vert_{\partial_\lambda\ra\partial_\lambda+ (z_u/2) \alpha\,\omega_\lambda}
=dx^\lambda \hat\nabla_\lambda u^\mu,
\eea
where $\tilde\nabla_\lambda u^\mu=\partial_\lambda u^\mu +\tilde\Gamma_{\lambda\rho}^\mu u^\rho$
and where $\hat\nabla_\lambda u^\nu$ is indeed Weyl covariant under Weyl gauge transformation (\ref{WGS})
as one can easily see by using that $\tilde\Gamma$ is Weyl gauge invariant.
From (\ref{gg})  the differential variation of the vector $d u^\mu\equiv dx^\lambda\, \partial_\lambda u^\mu$
is then
\bea\label{va}
d u^\mu=-dx^\lambda \big[ (1/2) z_u \,\alpha\, \omega_\lambda \, u^\mu +\tilde\Gamma_{\lambda\rho}^\mu \,u^\rho].
\eea
Using $(\tilde\nabla_\lambda\! +\!\alpha\,q\, \omega_\lambda) g_{\mu\nu}=0$, then
the change of the norm $d\vert u^\mu\vert^2=d (u^\mu u^\nu g_{\mu\nu})$ has  solution
\bea\label{norm}
\vert u\vert^2=\vert u_0\vert^2 \,e^{-\alpha (q+z_u) \, \int_\gamma \omega_\lambda dx^\lambda}.
\eea
%
Now the tangent space vector $u^a=e^a_\mu \, u^\mu$ has Weyl charge equal to $q/2+z_u/2$,
being given by the sum of charges of $e^a_\mu$ which is $q/2$ (half of that of $g_{\mu\nu}$)
and of $u^\mu$ ($z_u/2$). Then if $u^a$ of the tangent plane  is invariant, i.e. has Weyl charge zero, 
$(q+z_u)/2=0$, then from  (\ref{norm})   the norm of $\vert u^\mu \vert =\vert u_0\vert$
is thus invariant under parallel transport. 
This result can also be seen from an equivalent  formulation which is metric but has torsion
\cite{CDA}. The mathematics is  identical but in this case vectorial non-metricity
is  interpreted as vectorial torsion!

Therefore,  the norm of a vector is invariant and does not depend on the path history,
contrary to the long-held view, if  Weyl gauge-covariance is respected (as it should in a gauge
theory).
Hence, although Weyl geometry is historically called ``non-metric''
i.e.  $\tilde\nabla_\mu g_{\alpha\beta}\not=0$,
the  norm of a vector does not change under parallel transport\footnote{
  Given this result,  the term ``non-metric'' Weyl geometry
  is a misnomer and it strictly means  $\tilde\nabla_\mu g_{\alpha\beta}\not=0$.}. This is
immediately obvious in the Weyl gauge covariant (metric) formulation (of $\hat\nabla$)
where $\hat\nabla_\mu g_{\alpha\beta}=0$!

Regarding {\bf  b)} above, to have a second clock effect one needs a clock rate and this
requires in turn  the presence of a mass term  in the Lagrangian of the theory.
In the absence of a mass scale there is no clock rate and thus, there is  no second-clock effect!
One can define a clock rate by the  Hawking radiation, but that requires
the presence of a black hole of some mass, so we need again a mass scale in the action.
In the Weyl gauge invariant phase eq.(\ref{Sf0}), (\ref{inA}),  the Lagrangian has no mass scale,
 then a clock rate cannot be defined;  thus one cannot have a
 second clock effect in the symmetric phase\footnote{In \cite{Lasenby}
 an additional scalar field compensator  was introduced in this phase,
  however there is no need for this step -
 the  clock rate is defined by the  mass generated in the broken phase only, 
 without adding any  scalar field compensator,  see  Section~\ref{wgsb}.}.

In the broken phase of Weyl gauge symmetry of the action, 
we saw that the gauge field  $\w_\mu$  becomes  massive and thus can decouple below $m_\w$
leaving behind a broken phase of the theory in which one recovers Einstein
gravity and the (metric) Riemannian geometry. Now there  is also a clock rate
in the theory. Any non-metricity effects that one would still  claim,
such as second clock effect, are then   suppressed by $m_\w$ \cite{Ghilen0}.
To conclude, the mentioned criticism does not
apply to Weyl conformal geometry and its quadratic gravity action.

\subsection{The limit of vanishing Weyl current is  conformal gravity}\label{j}

Weyl quadratic gravity   has a conserved current. From (\ref{alt2}) by applying Riemannian
operator
$\nabla_\mu$ to the equation of motion of $\w_\mu$ one finds in $d=4$ 
\cite{Ghilen0} and \cite{SMW} (Appendix B) 
\medskip
\bea\label{jj}
J_\mu=-\frac{\alpha\,q}{2 \xi^2}\,\phi\, (\partial_\mu+\alpha\,q_\phi\,\, \w_\mu)\, \phi, \qquad
\nabla_\mu J^\mu\!=\!0,\qquad (q_\phi=-q/2).
\eea
$J_\mu$ can also be written as below,
using that $\phi^2=-\hat R$ and the charge $-q$ of $\hat R$:
\bea
J_\mu=
\frac{\alpha\, q}{4\xi^2}\,\, \hat \nabla_\mu \hat R.
\eea

Consider the limit this current vanishes $J_\mu=0$ which happens for
$\omega_\mu=2/(\alpha\,q) \, \partial_\mu \ln\phi$.
This means that the Weyl gauge boson is not dynamical because its field strength 
$F_{\mu\nu}=\partial_\mu\w_\nu-\partial_\nu\w_\mu$ vanishes $F_{\mu\nu}=0$
($F_{\mu\nu}$ is also  known as the length curvature tensor).
In this case Weyl geometry is ``integrable''  which is 
conformal to Riemannian geometry; the Weyl connection is  a conformally
transformed (via (\ref{WGS})) Levi-Civita connection (\ref{tildeGamma}).
Then  the above $\omega_\mu$ can  be integrated out and  action (\ref{alt2}),
with $F_{\mu\nu}=0$, becomes
\bea\label{lim}
S_{\bf w}=\int d^4x \sqrt{g} \,\Big\{
-\frac{1}{2\xi^2} \,\Big[\,\frac16\,\phi^2\,R+(\partial_\mu\phi)^2\Big]-\frac{1}{4! \xi^2}
\,\phi^4-\frac{1}{\eta^2}\, C_{\mu\nu\rho\sigma}^2\Big\}.
\eea

\medskip
We  recovered the conformal gravity action \cite{mannheim,Mannheim2}
with the benefit of an additional local Weyl invariant action of a
dilaton, which has a {\it geometric origin} in the $\hat R^2$ term,
hence it is not added ad-hoc. This is a certain advantage since
the dilaton in (\ref{lim}) generates the Planck scale and the Einstein
term when it acquires a vev $\phi\!\ra\! \langle\phi\rangle$ and
conformal symmetry is broken. The broken phase action is as in (\ref{EP}), (\ref{la})
with  $\omega_\mu$ dependence removed (formally $\omega_\mu\ra 0$).
The dilaton, as a dynamic degree of freedom,
also contributes to anomaly cancellation by mixing with the graviton,
to realise quantum conformal gravity \cite{Englert} (Section~\ref{wa}).

Briefly, conformal gravity is  just a particular limit of vanishing Weyl current of 
the Weyl quadratic  gravity of eq.(\ref{alt2}) \cite{SMW2}. This is
interesting, but the result is somewhat  expected  since gauging the conformal
group demands an integrable Weyl geometry \cite{Kaku} (obtained for $J_\mu=0$);
in such case
no gauge kinetic terms are allowed for either the special conformal symmetry
(not considered here) or for gauge dilatations (corresponding to $F_{\mu\nu}=0$).
This explains why conformal gravity is recovered in the limit taken here\footnote{
This is due to the fact that, ultimately, conformal gravity is not really a gauge theory of the
conformal group since the spectrum has no physical associated gauge bosons;
or it is a gauge theory just as much as
the electroweak  theory would be without kinetic terms for the
electroweak gauge bosons $Z$, $W^\pm$.}.
For a study of this class of theories see \cite{t4,t2,mannheim,t3}.
To conclude, action (\ref{inA}), (\ref{alt2}) is then more general.

\subsection{SM in Weyl conformal geometry}\label{2.3}

The discussion so far did not include matter fields.
For a realistic theory, consider then  embedding the SM in Weyl conformal geometry,
for all details see \cite{SMW}. This embedding is {\it natural}, because
the SM action is scale invariant for a vanishing Higgs mass which  in the SM is actually a
parameter\footnote{The Higgs mass is generated after spontaneous breaking of
Weyl gauge symmetry.}. Therefore,  the SM action  is made invariant under
Weyl gauge  transformation with an only minor update in the Higgs sector,
as we explain. The SM embedding is also {\it minimal} because no additional degrees of
freedom are needed beyond those of the SM and Weyl geometry. Let us detail.

Firstly, the SM gauge bosons action is 
\bea\label{sg}
S_g=-\sum_{\rm{groups}}\int d^4x \frac{\sg}{4} g^{\mu\rho} g^{\nu\sigma} F_{\mu\nu} F_{\rho\sigma},
\eea

\medskip\noindent
$F_{\mu\nu}$ involves the difference $\tilde\nabla_\mu A_\nu -\tilde\nabla_\nu A_\mu$, where
we used $F_{\mu\nu}$ and $A_\mu$ as generic notations for a SM gauge boson and
since $\tilde\nabla_\mu A_\nu=\partial_\mu A_\nu -\tilde\Gamma_{\mu\nu}^\rho A_\rho$, then
for a symmetric $\tilde\Gamma_{\mu\nu}^\rho=\tilde\Gamma_{\nu\mu}^\rho$ the
$\w_\mu$-dependence cancels out in the field strength.
The SM gauge bosons action is  independent of $\w_\mu$ and has the same expression
as in the Riemannian case. Thus there is no classical coupling of SM gauge bosons to
$\w_\mu$. The invariance of $S_g$ under (\ref{WGS}) is obvious, because
the field strength is invariant  in $d=4$, while $\sqrt{g}\, g^{\mu\rho}\, g^{\nu\sigma}$
is also invariant.

Consider next  the Dirac part of the SM action; we show below that
this action is independent of $\w_\mu$, too, even though the fermions have non-vanishing Weyl
charges, eq.(\ref{WGS}). The action can most
easily be obtained from the Riemannian one by a suitable
``covariantisation'' of the derivative $\partial_\mu\psi\ra(\partial_\mu +q_\psi \alpha \w_\mu)\psi$
acting on the fermions fields,  with respect to Weyl gauge symmetry.
The spin connection can also be
found by a  ``covariantisation'' of the  Riemannian spin connection $s_\mu^{ab}$
with respect to this symmetry
\medskip
\bea
\tilde s_\mu^{ab}=
s_\mu^{ab}\Big\vert_{\partial_\mu e_\nu^b\ra \big[\partial_\mu+
   \alpha\,(q/2)\,\w_\mu\big] e_\nu^b}
= s_\mu^{ab} + 
\, \alpha \,(q/2)\, (e_\mu^a e^{\nu b} -e_\mu^b e^{\nu a}) \,\w_\nu,
\eea

\medskip\noindent
where we used that $e_\nu^b$ has Weyl charge $q/2$, with the Riemannian spin connection
\bea
s_\mu^{ab}=-e^{\lambda \,b} (\partial_\mu e^a_\lambda-\Gamma_{\mu\lambda}^\nu\,e^a_\nu).
\eea
One can check that $\tilde s_\mu^{ab}$ is invariant under (\ref{WGS}).
The Dirac  action in Weyl geometry is then
 \medskip
\bea\label{Lf}
S_f\!\!\!& =&\!\!\!\int d^4x \,\frac12\, \sg\,\, \overline\psi \, i\,\gamma^a \, e^\mu_{\,a}\,
\,\Big[
(\partial_\mu +q_\psi\, \alpha\,\w_\mu) - i g\, \vec T \vec{A}_\mu - i\, Y g^\prime  B_\mu
+\frac12 \tilde s_\mu^{\,\,ab} \,\sigma_{ab} \Big]\,\psi+h.c.\qquad
\\
&=&\int d^4x \,\frac12\, \sg\,\, \overline\psi \, i\,\gamma^a \, e^\mu_{\,a}\,
\,\Big[
\partial_\mu - i g\, \vec T \vec{A}_\mu - i\, Y g^\prime  B_\mu
+\frac12  s_\mu^{\,\,ab} \,\sigma_{ab} \Big]\,\psi+h.c.\label{wf}
\eea

\medskip\noindent
with $q_\psi=(-3/4)\, q$ for $d=4$ and $\sigma_{ab}=(1/4)[\gamma_a,\gamma_b]$.
Eq.(\ref{wf}) is actually the Riemannian version of  Dirac action: what happened is that
in (\ref{Lf}) the dependence on $\w_\mu$ of the covariant derivative has cancelled the
$\w_\mu$-presence in $\tilde s_\mu^{ab}$ \cite{Kugo}\footnote{For a
review see Section 2.3 and Appendix~A
in \cite{SMW}.}. The action is easily verified to be  Weyl gauge invariant.
In $S_f$ the usual quantum numbers of the SM fermions apply (not shown),
$\vec T=\vec\sigma/2$,  with ($g$, $A_\mu$) and ($g^\prime$, $B_\mu$) the gauge couplings and fields
of $SU(2)_L$ and $U(1)_Y$, respectively.
The fermionic action is thus similar to that in Riemannian geometry
and there is no classical coupling of fermions to $\w_\mu$.

There is an interesting exception to this result. The symmetry of the theory and the
Coleman-Mandula theorem do not forbid a gauge kinetic mixing of the gauged dilatation
field $\w_\mu$ (of field strength $\hat F_{\mu\nu}$)  to
the $B_\mu$ gauge field of hypercharge (of field strength
$F_{y \,\mu\nu}$)
\be
\label{EPp}
S_m\!=
-\frac12\int d^4x\, \sqrt{g} \sin \chi\, \hat F_{\mu\nu}\,  F_y^{\mu\nu}.
\ee

\medskip\noindent
This kinetic mixing is removed by the transformation below,  to  new ('primed') fields
\bea\label{bwp}
 \w_\mu  =  \w'_\mu \sec\chi,
\qquad
B_\mu =  B'_\mu-\w'_\mu\,\tan\chi.
\eea

\medskip\noindent
In the new  basis ($B_\mu^\prime, \w_\mu^\prime$)
the gauge kinetic mixing is removed, at the expense of a small
change in the fermionic action (\ref{wf})
with substitution (\ref{bwp}) that induces a coupling of fermions to $\w_\mu^\prime$.
This coupling does not introduce a gauge anomaly because the coupling is proportional to
the hypercharge $Y$ which is anomaly-free for the SM spectrum. Therefore the
Weyl gauge boson $\w_\mu$ is both massive and anomaly-free\footnote{
This is similar to massive and
anomaly-free gauge bosons in interesting D-branes models \cite{int1,int2}.}.
The above  coupling vanishes in the absence of the mixing. 
Ultimately,
this result  says that  Weyl's initial  attempt to interpret $\w_\mu$ as the real photon
was not entirely wrong, since  $\w_\mu$ can mix with the hypercharge gauge field!
This  effect is however very small \cite{SMW}.

The Yukawa interactions  are as in the SM, ``upgraded'' to curved-space time:
\medskip
\bea\label{Y}
S_Y=\int d^4x \, \sqrt{g} \sum_{\psi=l,q} \Big[\overline \psi_{L} Y_\psi H \psi_{R}
+ \overline \psi_{L} Y'_\psi \tilde H \psi_{R}'\Big] + h.c.,
\eea

\medskip\noindent
where $H$ is the Higgs $SU(2)_L$ doublet and $\tilde H=i\sigma_2 H^\dagger$,
the sum is over leptons and quarks; $Y, Y'$ are the SM  Yukawa matrices.
This action  is Weyl gauge invariant: indeed,  the sum of the  Weyl charges
(which are real) vanishes for each term, see (\ref{WGS}).
Hence this part of the action is again like in the Riemannian space-time.

The SM Higgs sector  is the only SM sector that undergoes small changes
to make it Weyl gauge invariant. First,  the derivative
in the Higgs kinetic term is upgraded to a Weyl-covariant one and second,
there exists a non-minimal coupling of the Higgs to the Weyl scalar curvature.
Hence, the Higgs action is
 \medskip
  \be\label{higgsR}
S_H\!=\int d^4x\!\sg\,\Big\{\,
-\,\frac{\xi_h}{6}\,\vert H \vert^2 \hat R +\vert \tilde D_\mu H\,\vert^2
-{\lambda}\, \vert H\vert^4 
\Big\}.
\ee
The $SU(2)_L \times U(1)_Y \times D(1)$ derivative acting on $H$ is
$\tilde D_\mu H=\big[\partial_\mu - i \cA_\mu 
-(q/2) \q\,\w_\mu\big]\, H$,
where  $\cA_\mu=(g/2) \,\vec \sigma.\vec A_\mu +  (g^\prime/2)\, B_\mu$;\,
$\vec A_\mu$ is the $SU(2)_L$ gauge boson, $B_\mu$ is the $U(1)_Y$ boson.
The kinetic term of the higgs then generates a new coupling in the action of the Higgs
to the Weyl gauge boson, coming from
\medskip
\bea\label{higgsD}
\vert \tilde D_\mu H\vert^2=\big\vert \big[\partial_\mu - \alpha \,(q/2)
\,\, \w_\mu\big] H\big\vert^2 +H^\dagger \cA_\mu \cA^\mu H.
\eea

\medskip
The Higgs sector can be analysed in the unitary gauge of $U(1)_Y$ symmetry
where $H=(1/\sqrt{2})\, \hat h \, \zeta$, with $\zeta^T\!\equiv\!(0,1)$. 
As done earlier in the absence of SM,
replace in the action $\hat R^2\ra -2\phi^2 \hat R -\phi^4$, to
linearise the quadratic term $\hat R^2$ in the Lagrangian.
The  non-minimal coupling then changes
$-(1/6)\,\xi_h \vert H\vert^2 \hat R\ra
- (1/12) \,\big[(1/\xi^2) \phi^2+\xi_h\,h^2\big] \hat R$.
Note the first term $(1/\xi^2) \phi^2 \hat R$ has actually a
non-perturbative coupling, since $\xi \ll 1$. 
This will not affect our calculation. Eq.(\ref{la}) for the Planck scale is
now modified into
\bea\label{47}
M_p^2\equiv \frac{1}{6\,\xi^2}
\Big(\langle\phi^2\rangle+\xi^2 \xi_h \langle \,\hat h^2\rangle\Big)
\eea
%
The ``radial direction'' in the field space of $\phi$ and $\hat h$
that generates now  the Planck scale is eaten by $\w_\mu$
which becomes massive, while the ``angular combination'' of $\phi$ and $\hat h$ is now
the (real) neutral scalar higgs, denoted $\sigma$ below; this is obtained after a redefinition
\bea\label{hh}
\hat h=M_p \sqrt{6}\sinh(\sigma/M_p\sqrt{6})
\eea
%
necessary to  ensure a canonical kinetic term  $(1/2) (\partial_\mu \sigma)^2$ for  $\sigma$.
The scalar potential becomes
\bea\label{h1}
V&=&\frac{1}{4!} \Big[ 6 \lambda \hat h^4+ \xi^2 (6 M_p^2-\xi_h \hat h^2)^2\Big].
\nonumber\\[5pt]
&=&
\frac{3}{2} \,M_p^4\, 
\Big\{{6 \lambda}\, \sinh^4 \frac{\sigma}{M_p\sqrt 6}
+
\xi^2\Big(1- \xi_h \sinh^2\frac{\sigma}{M_p\sqrt 6}\Big)^2\Big\}\label{h2}
\eea

\medskip\noindent
The higgs field  acquires  new couplings $\Delta L_h$
from the covariant derivative and, when present, also from the gauge kinetic mixing, which
are
\medskip
\bea\label{deltah}
\Delta S_h ={(1/8)}\int d^4x \sqrt{g}\, \sigma^2 \w_\mu^\prime \w^{\prime \mu}\,
\big[\alpha^2\,q^2\,\sec^2\chi+g^{\prime 2}\tan^2\chi\big]
\eea

\medskip\noindent
Here $\chi=0$ corresponds to the absence of a gauge kinetic mixing, when the
second term in $\Delta S_h$ vanishes. Constraints on $\alpha$ are very weak
so one can have $\alpha\sim \cO(1)$ then $m_\w$ is close
to the Planck scale while tuning it to ultraweak values $\alpha\ll 1$ then
$m_\w$ may become of the order of TeV scale. This may be even lower, as discussed
in Section~\ref{wgsb}.

To conclude, the total action of SM and  gravity in Weyl conformal geometry, is then
\medskip
\bea\label{SMW}
S=S_{\bf w}+S_g+S_m+S_f+S_Y+S_H.
\eea

\medskip\noindent
 $S_{\bf w}$ is given in  eq.(\ref{inA}), with the remaining
terms shown in eq.(\ref{sg}), (\ref{wf}), (\ref{EPp}), (\ref{Y}), (\ref{higgsR}).
This action can be used in applications, for some implications
see \cite{SMW,SMW2}.

\subsection{Starobinsky-Higgs inflation}

An immediate application of    SM  action in
conformal  geometry is inflation \cite{WI1,WI3}; here we follow \cite{WI1} (see
also \cite{WI2,SMW}). For large field values, the higgs field $\sigma$ plays
the role of inflaton. First, note that large field
values, near or above the Planck scale,  are actually natural in Weyl geometry.
This is because, as we have seen, the Planck scale is simply a scale where
Weyl gauge symmetry is spontaneously broken, i.e. it is just a phase transition scale.

For successful inflation from potential (\ref{h1}),
it  is necessary that the higgs self-coupling be very small
$\lambda\ll 1$, so that  the second term in (\ref{h2}) can dominate - this term
arises from the   $\phi^4$ term generated by the ``scalaron''
(or the would-be-Goldstone/dilaton) field that linearised the $\hat R^2$ term
in action $S_{\bf w}$. As a result, not surprisingly, the predictions from inflation
will be similar to those in  Starobinsky inflation \cite{Starobinsky}.
We thus have a gauge invariant version of the Starobinsky model of inflation.
Using potential (\ref{h2}), a clear prediction from our Starobinsky-Higgs
inflation in Weyl geometry  is found \cite{WI1}
\bea\label{rel}
r=3\, (1-n_s)^2-\frac{16}{3} \xi_h^2+\cO(\xi_h^3)
\eea

\noindent
where $r$ is the tensor-to-scalar ratio and $n_s$ is the scalar spectral index;
$\xi_h$ is the Higgs non-minimal coupling, eq.(\ref{higgsR}).
Ignoring the  $\xi_h$-dependent terms, relation (\ref{rel}) is similar to that
in Starobinsky inflation which is saturated for $\xi_h=0$ (when
Higgs  decouples from the scalar curvature $\hat R$ of  Weyl geometry).
Then for a given $n_s$, the effect of  $\xi_h$ is  to reduce  the value of $r$
from that in Starobinsky inflation.  The  numerical results give that for $N=60$ efolds and 
$n_s=0.9670\pm 0.0037$ at $68\%$ CL (TT, TE, EE+low E + lensing + BK14 + BAO) \cite{planck2018}
the following values for $r$  \cite{WI1,WI2,WI3}
\be\label{pi}
0.00227\leq r\leq 0.00303, \quad (n_s\,\,\,\textrm{at}\,\,\,95\% \,\,\textrm{CL}).
\ee

\noindent
The  upper limit corresponds to  Starobinsky inflation for $N=60$ e-folds.
Remarkably, the  results quoted above from \cite{WI1} and  obtained from
the scalar potential $V(\sigma)$  of (\ref{h2})
agree both  numerically and  analytically to  those in \cite{WI3}
obtained  by a different  method that used a two-field analysis of inflation.
This is a good check of prediction (\ref{pi}).

The discussion so far ignored the effect on inflation
of the Weyl term $(1/\eta^2)\,C_{\mu\nu\rho\sigma}^2$ present in the action of
Weyl quadratic gravity. This effect  is small
as long as $\eta$ is not too far from unity.
As expected, the relative variation of  $r$ due to the Weyl term depends on the
relative  value of the couplings $\xi$ versus $\eta$,
and we have $\delta r/r\approx 4\, \xi^2/\eta^2$ (assuming $\xi/\eta\ll 1$) \cite{WI1}.
This change is minimised for $\eta\sim 1$, when the mass $\eta M_p$  of the
spin-two state associated to $C_{\mu\nu\rho\sigma}^2$ is near the Planck
scale $M_p$ (for  details see \cite{WI1}).

The small values of $r$ shown above may be reached by the next
generation of CMB experiments CMB-S4 \cite{CMB1,CMB2}, LiteBIRD
\cite{litebird,CMB3}, PICO \cite{CMB4},
PIXIE \cite{Pixie}, which have sensitivity to $r$ up to  $0.0005$.
Such sensitivity can test this  inflation model and even distinguish it
from the Starobinsky model for $\xi_h\sim 10^{-2}$ when the effect of
$\xi_h$ on the curve $r(n_s)$ becomes apparent, see the plots $r(n_s)$
in \cite{WI1}.
Somewhat similar  values of $r$ are also found in other models with
Weyl gauge symmetry \cite{WI2} based on a Palatini approach to  the
quadratic gravity action in eq.(\ref{inA});  these models do not respect
relation (\ref{rel}) and the slope of the curve $r(n_s)$ is different,
with mildly larger $r\sim 10^{-2}$ due to their different vectorial non-metricity.
To conclude,  assuming the inflation scenario describes physical reality,
then Weyl quadratic gravity  leads to a clear, narrow prediction
for $r$ that will  soon  be tested experimentally.

\subsection{Origin of masses,  hierarchies of scales}

The  Weyl gauge theory in (\ref{SMW})  has the following
parameters, in addition to those in SM:

$\xi<1$: the dimensionless coupling of  $\hat R^2$,

$\alpha<1$: the dimensionless Weyl gauge coupling,

$\eta<1$: the dimensionless coupling of the Weyl tensor-squared term in the action.

$\langle\phi\rangle$:  vev of $\phi$ (would-be-Goldstone of Weyl gauge symmetry
or ``dilaton'' $\ln\phi$).

\noindent
These parameters enabled us to fix (classically) the  following couplings/scales:

1) the cosmological constant $\Lambda$, by fixing $\langle\phi\rangle$, eq.(\ref{la}).

2) the  Planck  scale $M_p$ by fixing $\xi$, see eq.(\ref{la}), (\ref{47}).

3) the mass of $\omega_\mu$,  which depends on $\alpha$,

4) $\eta$ is fixed to ensure a large mass ($\eta\, M_p$) of the massive
spin-2 state induced by $C_{\mu\nu\rho\sigma}^2$ term in eq.(\ref{EP}) \cite{SMW,LAG}.
For natural values of  $\eta\sim \cO(1)$, that we assume in this work and in (\ref{EP}),
this state naturally decouples near Planck scale.  Classically one may want to
remove  $C_{\mu\nu\rho\sigma}^2$ from the action (formally $\eta\ra \infty$);
we address later what happens at quantum level.

There is one more parameter, the non-minimal coupling of the Higgs $\xi_h<1$ that
helps fixing the hierarchy between the Higgs vev $\langle\sigma\rangle$
and that of the ``dilaton'' $\langle\phi\rangle$ \cite{SMW}.

To fix the values of the above scales there is no fine-tuning 
understood as a special cancellation in a sum of large numbers.
As mentioned, all these mass scales, $M_p$, $\Lambda$, $m_\w$ have ultimately
a {\it  geometric origin}; this is because their values are proportional to the vev of $\phi$
(which was not added ad-hoc as in other theories), but it was
``extracted'' from the quadratic term in the action, $\hat R^2$, which
is part of geometry. This is an elegant feature of the theory.
Another important advantage compared to other theories is that,
with $\phi$  eaten by $\omega_\mu$ (which thus becomes
massive), there is {\it no need to stabilize} the vev of $\phi$!

Given the coupling $\w_\mu \w^\mu \sigma^2$ (\ref{deltah}),
then in the symmetric phase of the action e.g. in the early Universe, when $\w_\mu$ is 
massless, the neutral Higgs $\sigma$ can be generated via vector boson fusion.
If so, ultimately the the Higgs boson field  itself has a geometric origin
(just like  $\omega_\mu$)  in the underlying Weyl geometry. Since the
SM masses are generated by the Higgs field, one may say that ultimately
all masses are related to the underlying geometry. 
Further,  if the mass of $\w_\mu$ is very light, one cannot exclude a scenario
in which $\sigma+\sigma \ra \w_\mu + \w_\mu$ takes place. Since  $\w_\mu$ 
contributes to the Weyl scalar curvature of the underlying  geometry,
this process modifies locally the space-time curvature. Such process
is interesting and can  have other implications, 
depending on the value of  $\alpha$.

At the quantum level,  the mass of the Higgs, $m_\sigma$, could acquire large quantum
corrections,  like in the SM, to which one should also add those due to
the coupling  $\w_\mu \w^\mu \sigma^2$. This situation can be changed by the
Weyl gauge symmetry, if its breaking scale identified with the mass of the Weyl boson
$m_\w\propto \alpha M_p$ is very
low, possible for ultraweak $\alpha\ll 1$. As discussed earlier, the  current bound 
is $m_\w\sim 1$ TeV but may be even lower. Then  if the breaking scale of Weyl gauge symmetry
is near the TeV scale and with the  quantum corrections to the Higgs mass being quadratic
in the scale of ``new physics'', then
\bea
\delta m_\sigma^2\sim m_\w^2.
\eea
For a light $m_\w$ this correction is naturally small.
Above $m_\w$,  Weyl gauge symmetry is restored together with
its protection to $m_\sigma$, since no mass counterterm to $\sigma$ is allowed by this 
symmetry of  the action. In this way,  Weyl gauge symmetry could provide a solution
to the mass hierarchy problem. This issue deserves a careful study at one-loop level.

\subsection{Weyl anomaly: Riemann vs Weyl geometry}\label{wa}

Let us now discuss Weyl anomaly \cite{Duff,Duff2,Deser,Duff3,Deser1976}
in Riemann versus Weyl geometry, following \cite{DG1}.

Quantum corrections are in general divergent. Therefore,  to begin with, one is required
to perform an analytical continuation of the action  to $d=4-2\epsilon$ dimensions
so that it makes mathematical
sense.  Quite often such analytical continuation  breaks  the symmetry which then
does not survive at the quantum level and anomalies may  be present. But anomalies
are more than a regularisation-related issue and may e.g. signal the missing of a
(dynamical) degree of freedom.
With scale symmetry as a gauge symmetry, an immediate question arises:
Weyl symmetry (i.e. local scale invariance but no $\w_\mu$)
is in general anomalous \cite{Duff,Duff2,Duff3,Deser1976},
then how is Weyl anomaly reconciled with the Weyl {\it gauged} scale symmetry?
In the case of a gauge theory as here, it is crucial to ensure that this
symmetry survives at the quantum level and is anomaly-free, so that
this gauge theory is quantum consistent, as a candidate for a quantum gravity theory.

Here we review comparatively the issue of Weyl anomaly
in Riemann versus Weyl  geometry -based theories of gravity.
We show why Weyl geometry as a gauge theory of dilatations naturally
avoids  Weyl anomaly (of both Weyl and Euler terms), to
become a quantum consistent gauge symmetry and how this anomaly is ``recovered''
in the  broken phase.

\bigskip
\noindent
{\bf $\bullet$ Riemannian geometry:   }

\bigskip\noindent
Weyl anomaly  \cite{Duff,Duff2,Duff3,Deser1976,Deser} arises
partly because the regularisation of the quantum corrections due to some massless fields
does not respect the classical Weyl symmetry of their action.
Dimensional regularisation (DR), through the presence of the subtraction scale $\mu$,
breaks explicitly this symmetry. One can avoid this situation with a Weyl
invariant regularisation introduced in
\cite{Englert}\footnote{Keeping global or local scale symmetry at the quantum level was
studied up to 3 loops \cite{Englert,Misha1,Misha2,Misha2s,Misha3,Tamarit,dg1,dg2,dg3,dg7,3loop}.}.
In this way Weyl anomaly can be avoided as done in  conformal gravity  and the symmetry is
maintained at quantum level to give a quantum conformal gravity theory \cite{Englert}.
However,  Weyl anomaly is more than just
a regularisation problem:  the Euler-Gauss-Bonnet term in the action, which is a total derivative
in $d=4$, brings an anomaly which is $\mu$-independent and hence cannot be avoided by the same
approach. 

Let us first review the usual Weyl anomaly in Riemannian geometry
\cite{Duff,Duff2,Duff3,Deser,Deser1976}
that emerges at quantum level in a theory with classical {\it Weyl symmetry}. 
By {\it Weyl symmetry} we mean the (local) invariance of the action  under a transformation 
of the metric $g_{\mu\nu}$ as in (\ref{WGS}) and, if present,  of its scalar(s) $\phi$ and
fermion(s) $\psi$, but no $\w_\mu$ is present; in $d$ dimensions the transformation is
\medskip
\bea\label{WGi}
g_{\mu\nu}'=\Sigma^q g_{\mu\nu}, \quad
\phi'=\Sigma^{q_\phi} \phi,
\quad
\psi'=\Sigma^{q_\psi}\psi,
\quad
q_\phi=-\frac{q}{4}\,(d-2),
\quad
q_\psi=-\frac{q}{4}\, (d-1).
\eea

\medskip
Next, consider a  Weyl-invariant action of some (unspecified) massless matter
fields that
interact only gravitationally with an external gravitational field and an external spin-1
gauge field. It is assumed that there are no self-interactions of these matter fields.
Then at one-loop
 a gravitational effective action is induced by the (divergent) quantum corrections from
these matter fields. In the DR scheme this action has the form
\medskip
\bea\label{Wd}
W_d=\frac{1}{d-4} \int d^d x \sqrt{g} \, A(d), \qquad d=4-2\epsilon.
\eea

\medskip\noindent
where $A(d)$ if a function of the metric and its derivatives; it contains higher derivative
operators such as those below, in a combination such that 
 $W_d$ is Weyl invariant \cite{Buch}:
\medskip
\bea\label{ops}
R \,X \,R,\quad
R_{\mu\nu}\, X\, R^{\mu\nu},\quad
R_{\mu\nu\rho\sigma}\, X\, R^{\mu\nu\rho\sigma},\quad
F_{\mu\nu} \,X\, F^{\mu\nu},\,\,
C_{\alpha\beta\gamma\delta} \,X\, C^{\alpha\beta\gamma\delta}, \,\,
\,X\, \equiv\Box^{(d-4)/2}.
\eea

\medskip\noindent
Here  $\Box=\nabla^\mu \nabla_\mu$ in a  Riemannian
notation and $R$, $R_{\mu\nu}$, $R_{\mu\nu\rho\sigma}$ are
scalar and tensor curvatures of Riemannian geometry. $W_d$ has a pole $1/(d-4)$ seen when one expands
$\Box^{(d-4)/2}=1+(d-4)/2\ln\Box$. To cancel this pole the counterterm
\medskip
\bea\label{wc}
W_c=-\frac{\mu^{d-4}}{d-4}\int d^dx \sqrt{g}\,
\big( b \, C_{\mu\nu\rho\sigma}^2+ b^\prime G+ c \, F_{\mu\nu}^2\big).
\eea

\medskip\noindent
is added to the action, in an independent ``basis'' of Weyl invariant operators in $d=4$.
Here $b, b^\prime, c$ are some  constants that depend only on the matter field
content considered that runs in the loop, and
\bea
G\equiv R_{\mu\nu\rho\sigma}^2-4 R_{\mu\nu}^2 +R^2.
\eea

\medskip\noindent
In the limit $d\ra 4$, $G$ becomes a total derivative, the so-called
``Euler term''. $F_{\mu\nu}$ is now the field strength
of the external spin-1 gauge field mentioned.

The trace of the energy-momentum tensor is 
\bea\label{tmunu}
T^\mu_\mu=\frac{-2}{\sqrt{g}}g_{\mu\nu}\frac{\delta (W_d+W_c)}{\delta g_{\mu\nu}}\Big\vert_{d\ra 4}
=\frac{-2}{\sqrt{g}} g_{\mu\nu} \frac{\delta  W_c}{\delta g_{\mu\nu}}\Big\vert_{d\ra 4}.
\eea

\medskip\noindent
This trace is evaluated using that \cite{Duff}
\medskip
\bea\label{eb}
\frac{2}{\sqrt g} g_{\mu\nu}\frac{\delta}{\delta g_{\mu\nu}}
\int d^d x \sqrt{g} \, C_{\mu\nu\rho\sigma}^2
&=&(d-4) \big( C_{\mu\nu\rho\sigma}^2
+ \frac23 \,\Box R\big),
\\
\frac{2}{\sqrt g} g_{\mu\nu}\frac{\delta}{\delta g_{\mu\nu}}\int d^d x \sqrt{g}\, G&=&(d-4)\, G,\label{eb2}
\\
\frac{2}{\sqrt g} g_{\mu\nu}\frac{\delta}{\delta g_{\mu\nu}}\int d^d x \sqrt{g} \, F_{\mu\nu}^2
&=&(d-4)\, F_{\mu\nu}^2.
\label{eb3}
\eea

\medskip\noindent
With these relations\footnote{We outline here a derivation of eq.(\ref{eb}) \cite{Buch}; we have
  $\sqrt{g^\prime} \, K (g_{\mu\nu}') \!=\! \sqrt{g} \, K (g_{\mu\nu})\, \Sigma^{q\,(d-4)/2}$,
  where we denoted $K=C_{\mu\nu\rho\sigma}^2$.
Next, for  a functional $A(g'_{\mu\nu})$  under transformation (\ref{WGi})
$\delta A/\delta \ln\Sigma^q=(\delta A/\delta g'_{\mu\nu}) (\delta g'_{\mu\nu}/\delta\ln \Sigma^q)=
(\delta A/\delta g'_{\mu\nu})  g_{\mu\nu} \Sigma^q=g'_{\mu\nu}\delta A/\delta g'_{\mu\nu}.
$ 
For an infinitesimal transformation (\ref{WGi}) 
($\Sigma\!\ra\! 1$, $g'_{\mu\nu}\!\ra\!  g_{\mu\nu}$)
 and with a notation  $\cI(K)\equiv\int d^dx \sqrt{g'} K(g')$ we  find  
$(2/\sqrt{g'}) \,\, g'_{\mu\nu}(\delta/\delta g'_{\mu\nu}) \cI (K)
=(2/\sqrt{g'})\,\, \delta \cI(K)/\delta \ln\Sigma^q\vert_{\Sigma\ra 1}
= (d-4) K$, as shown in  (\ref{eb}). This misses the
 $\Box R$ term due to an ambiguity in local conformal case
\cite{Duff,Asorey}, but it is easily accounted for \cite{Buch} (section 17.2.2).}
used in (\ref{tmunu}), the pole in  $W_c$  is cancelled  by the factor $(d-4)$
in these equations; one then takes the limit $d\ra 4$, then
\medskip
\bea\label{anomaly0}
T_\mu^\mu=
 b \, [\,C_{\mu\nu\rho\sigma}^2+ (2/3) \Box R\,]+ b' G +c H.
 \eea

 \medskip\noindent
 Since this trace is non-vanishing, Weyl symmetry is thus anomalous.

 Soon after this calculation, an attempt to construct an anomaly-free
 quantum conformal gravity action was made in \cite{Englert}.
 Let us review this here.
 The authors considered the massless QED corrections to the conformal gravity action defined by
 $C_{\mu\nu\rho\sigma}^2$ term,  extended by a conformally-coupled scalar field $\phi$ (dilaton) of action
 \medskip
 \bea\label{R}
W_\phi=-\int d^dx \sqrt{g}\,\frac{1}{12}\Big(\phi^2\,R
+\frac{4\,(d-1)}{d-2} \,g^{\mu\nu}  \partial_\mu\phi\partial_\nu\phi\Big).
\eea

\medskip\noindent
Since $\phi$ transforms as in (\ref{WGi}), then
$\ln\phi\ra\ln\phi+q_\phi\ln\Sigma$ so $\ln\phi$ transforms with a shift and thus plays the role
of a dilaton/Goldstone field.
The calculation of the anomaly associated with $C_{\mu\nu\rho\sigma}^2$
proceeds as earlier, except that  the
subtraction scale $\mu$  in the counterterm $W_c$ is now replaced by $\phi$
to maintain Weyl symmetry in $d$ dimensions \cite{Englert}
\medskip
\bea\label{weyl}
W_c =-\frac{b}{d-4}\int d^d x \sqrt{g}\, \,\phi^{2(d-4)/(d-2)} \,C_{\mu\nu\rho\sigma}^2.
\eea

\medskip\noindent
Now the pole in $W_d$ of (\ref{Wd})  due to $C_{\mu\nu\rho\sigma}^2$ is cancelled as before by the pole
in $W_c$ while $\mu$ is   generated spontaneously
when  $\phi$ acquires a vev, $\mu\! \sim\! \langle\phi\rangle$.
Hence a Weyl-invariant regularisation was enforced at the ``cost'' of an {\it additional, dynamical
field} $\phi$ in the theory.

Since $W_c$ now respects the Weyl symmetry, the variation of $W_c+W_d$ under (\ref{WGi}) is now
vanishing; hence there is no contribution from $C_{\mu\nu\rho\sigma}^2$ to the trace of energy
momentum tensor and there is no anomaly from this term. The absence of an anomaly
in this case gave rise to different interpretations
\cite{Duff}. Here the absence of an anomaly  is due simply to the fact that the
theory has an additional dynamical degree of freedom (the dilaton $\ln\phi$)
which mixes with the graviton and contributes
to anomaly cancellation, see e.g. figure 3 in  \cite{Englert}.
Its decoupling restores the anomaly contribution to $T_\mu^\mu$.
In other words, Weyl anomaly simply signals the  missing or decoupling of a
dynamical degree of freedom that would otherwise enable the symmetry at  quantum level.

With (\ref{weyl}) one obtains a renormalised action $W_r=W_d+W_c$ plus a Weyl-invariant $W_\phi$
\medskip
\bea\label{1L}
W_{r}=\frac{b}{2}\int d^4x \sqrt{g}\,\,\,
C_{\mu\nu\rho\sigma} \ln \Big(\frac{\Box}{\phi^2}\Big)\, C^{\mu\nu\rho\sigma}.
\eea

\medskip\noindent
Here $\ln\Box$ ($\ln\phi^2$)
arise from $W_d$ ($W_c$) respectively, after the cancellation of the poles and after expanding
$\Box^{-\epsilon}=1-\epsilon\ln\Box$ and $(\phi^2)^{-\epsilon}=1-\epsilon\ln\phi^2$, which when
combined  generated $\ln(\Box/\phi^2)$ seen in $W_r$.
$W_r$ is then  Weyl invariant under an infinitesimal transformation (\ref{WGi}) in $d=4$.
This is so since we have $\sqrt{g^\prime}\, C_{\mu\nu\rho\sigma}^{\prime 2}=\sqrt{g}\, C_{\mu\nu\rho\sigma}^2$ but
$\Box^\prime =\Sigma^{-q} (\Box+f(\ln\Sigma))$ where $f(\ln\Sigma)$ depends on derivatives and is
neglected for an infinitesimal variation in (\ref{WGi}) when $g_{\mu\nu}\ra g_{\mu\nu}^\prime$,
$\Sigma\ra 1$. Then  $\ln(\Box/\phi^2)$ is Weyl invariant and so is the whole action $W_r$,
hence there is no anomaly, consequence of the presence of the additional dynamical field
($\phi$) \cite{Englert}.

When $\phi$ acquires a vev and decouples then Weyl anomaly is recovered. Let us detail.
By Taylor expanding $W_r$ about $\phi=\langle\phi\rangle+\delta\phi$ where $\delta\phi$ are
small fluctuations
and when these decouple ($\delta\phi\ll \langle\phi\rangle$),  one obtains 
\medskip
\bea\label{anomaly}
W_{r}=\frac{b}{2}\int d^4x \sqrt{g}\,C_{\mu\nu\rho\sigma}
\ln \Big(\frac{\Box}{\mu^2}\Big)\, C^{\mu\nu\rho\sigma},
\eea

\medskip\noindent
where we denoted $\mu\equiv\langle\phi\rangle$.
This result is identical to that one would have obtained
by the usual DR scheme, using  $\mu$ as regulator in $W_c$ of  (\ref{weyl}), as it has already
been seen for  the first term in eq.(\ref{wc}).
Then from (\ref{anomaly}) one has
\medskip
\bea T_\mu^\mu = - \frac{2}{\sqrt{g'}}\,\,  g'_{\mu\nu}\frac{\delta W_r}{\delta g'_{\mu\nu}}
= - \frac{2}{\sqrt{g'}} \,\, \frac{\delta W_r}{\delta \ln\Sigma^q}\Big\vert_{\Sigma\ra 1}
= b \, C_{\mu\nu\rho\sigma}^2.
\eea

\medskip\noindent
This is exactly the anomaly due to $C_{\mu\nu\rho\sigma}^2$ in (\ref{anomaly0}),
 re-derived  using   non-local  $\ln\Box$ term \cite{Buch}.

 Briefly, the Weyl-tensor-squared action is anomaly-free in the presence of a dilaton action and
 a Weyl-invariant regularisation enforced by the dilaton. The anomaly is ``restored'' in the decoupling
 limit of this field. However, this approach does not work if we add the Euler term to the action:
 unlike $C_{\mu\nu\rho\sigma}^2$, the Euler  term is {\it not} Weyl-covariant
 in $d$ dimensions and hence it cannot be made  Weyl invariant in $d$ dimensions\footnote{
 This is where Weyl geometry differs.} with the
 help of a Weyl-invariant regularisation as in eq.(\ref{weyl});
 this is because it is actually $\mu$-independent, type A anomaly \cite{Deser}.
  This ends our Weyl-anomaly review in the Riemannian case. See \cite{DG1} for more details.

\vspace{0.8cm}\noindent
 {\bf $\bullet$ Weyl conformal geometry:\,\,}

\bigskip\noindent
What is the situation for the action derived in Weyl geometry, with Weyl gauge symmetry?
What changes in Weyl geometry in an analysis similar to that above is that
the  Euler term is Weyl gauge covariant, eq.(\ref{WGS3}). Hence $\hat G$
can be  made Weyl gauge invariant by a suitable analytical continuation, as we saw
above for   $C_{\mu\nu\rho\sigma}^2$ in (\ref{weyl}).
Note the Weyl term is  Weyl-covariant in both Riemannian and
Weyl geometry since $\hat C_{\mu\nu\rho\sigma}^2=C_{\mu\nu\rho\sigma}^2$, eq.(\ref{ccc})).

Consider then the case of the general quadratic action in Weyl conformal geometry in which case
the symmetry is that of (\ref{WGS}) that extends Weyl symmetry of (\ref{WGi}).
The action is in eq.(\ref{Sf}) which we re-write below for convenience
\medskip
\bea\label{Sff}
S_{\bf w}=\int d^d x  \sqrt{g}\,\Big\{ a_0\hat R^2+ b_0 \hat F_{\mu\nu}^2+
c_0 \hat C_{\mu\nu\rho\sigma}^2 +d_0 \,\hat G\Big\} \, ({\hat R}^2)^{(d-4)/4}.
\eea

\medskip\noindent
This is already analytically continued to $d$ dimensions and is Weyl gauge invariant.

We can  consider the quantum  corrections to $S_{\bf w}$ from the
gravitational interactions of the SM states, using the action of SM in Weyl geometry \cite{SMW}.
For example, the Higgs (massless)  field had such interactions
\medskip
\bea\label{wh}
W_h\!=\!\int\! d^4 x\sqrt{g}\,\Big\{\frac12\,\hat \nabla_\mu h \hat\nabla^\mu h
-\frac{1}{12}\xi_h \, h^2 \hat R\Big\},\quad
 \hat\nabla_\mu h=\big[\partial_\mu  +\alpha \,q_h \w_\mu \big]\,h, \quad q_h= \frac{-q}{4} (d-2).
 \eea

\medskip\noindent
where $W_h$ is Weyl gauge invariant. One can also consider the gravitational
contributions of the SM gauge fields or fermionic
fields, since their action  is also Weyl gauge invariant, as we saw earlier.
The analysis is similar, the only difference is in the beta functions i.e.  the
coefficients of the quadratic terms in $S_{\bf w}$ that depend on the matter
fields   present in the loops of quantum corrections.

At the quantum level, the divergent vacuum action $W_d$ is similar to that in (\ref{Wd}), (\ref{ops})
\medskip
\bea\label{Wdp}
W_d=\frac{1}{d-4} \int d^d x \sqrt{g} \, A(d), 
\eea
but now
$A(d)$ is  a combination of the Weyl-covariant operators as seen in (\ref{WGS3}),(\ref{wq})
\medskip
\bea\label{nablas}
\hat R\, ( \hat \Box)^{(d-4)/2} \hat R,
\qquad
\hat R_{\mu\nu}\, ( \hat\Box)^{(d-4)/2} \hat R^{\mu\nu},
\quad
\hat R_{\mu\nu\rho\sigma}\, ( \hat \Box)^{(d-4)/2} \hat R^{\mu\nu\rho\sigma},\,\,\,
\textrm{etc.}, \quad \hat\Box\equiv \hat\nabla_\mu \hat\nabla^\mu,
\eea 

\medskip\noindent
and where   $\hat\nabla_\mu g_{\alpha\beta}=0$.
Therefore eq.(\ref{nablas}) is a Weyl-covariantised version of eq.(\ref{ops}).
The associated simple poles $1/(d-4)$  in $W_d$ can be cancelled by a counterterm $W_c$ that is 
Weyl gauge invariant in $d$ dimensions and has a general structure similar
to (\ref{weyl})\footnote{Equivalently, we could use the dilaton field as a regulator,
like in eq.(\ref{weyl}),  see later.}
\bea\label{op}
W_c=- \frac{1}{d-4}\int d^d x \sqrt{g}\,\Big\{a_1\hat R^2+ b_1 \hat F_{\mu\nu}^2+
c_1 \hat C_{\mu\nu\rho\sigma}^2 +d_1 \hat G\Big\}\, (\hat R^2)^{(d-4)/4}
\eea

\medskip\noindent
The vacuum part of the renormalised gravitational action is then
\bea
W_r=S_{\rm w}+W_d+W_c
\eea
Using  $W_d$  with eq.(\ref{nablas})
and $W_c$ of eq.(\ref{op}) and  $\Box^{(d-4)/2}=1+(d-4)/2\ln\Box+...$, then
\medskip
\bea\label{Wr}
W_r &= &
\int d^4 x \,\sqrt{g}\,\, \Big\{\hat R \,
\Big[a_0+\frac{a_1}{2} \ln\frac{\widehat\Box}{\vert \hat R\vert}\Big]\, \hat R
+  \hat F_{\mu\nu} \,\Big[b_0+\frac{b_1}{2} \ln\frac{\widehat\Box}{\vert \hat R\vert}\Big]
\,\hat F^{\mu\nu}
\nonumber\\[0pt]
&+&
\hat C_{\mu\nu\rho\sigma} \, \Big[c_0+\frac{c_1}{2}
\ln\frac{\widehat\Box}{\vert\hat R\vert}\Big] \hat C^{\mu\nu\rho\sigma}
+
\Big[\hat R_{\mu\nu\rho\sigma}
\, Z\,
\hat R^{\rho\sigma\mu\nu}
-4 \hat R_{\mu\nu} \,Z\,
\,\hat R^{\nu\mu}
+ \hat R
\,Z\,\hat R\Big] \Big\},
\eea

\medskip\noindent
where the last square bracket is due to the Euler term $\hat G$ with
\bea
Z\equiv d_0+\frac{d_1}{2} \ln\frac{\widehat \Box}{\vert \hat R\vert}.
\eea

\medskip\noindent 
$W_r$  includes UV non-local terms $\ln\widehat\Box$,  where 
$\widehat{\Box}=\hat\nabla_\mu \hat\nabla^\mu$ where $\hat\nabla$ is the Weyl gauge covariant
derivative. Each of the terms in (\ref{Wr}) are  Weyl gauge invariant,
because $\hat \Box$ and $\hat R$ are Weyl covariant
with the same charge ($-q$), hence their ratio is invariant and then the whole expression is invariant,
too. In conclusion we have an action which  has {\it manifest Weyl gauge symmetry at one-loop level,}
which indicates that this theory is Weyl-anomaly-free\footnote{
As it is usual  with gauge symmetries,
the absence of an anomaly here should remain true to all orders.}. This result is important
and specific to Weyl geometry. In this case the
trace of energy-momentum tensor is cancelled by the divergence of the 
Weyl current  (\ref{jj}), $J_\mu\sim \hat\nabla_\mu \phi^2$ 
\cite{DG1}
\bea
T_\mu^\mu -\frac{1}{\alpha'} \hat\nabla_\mu J^\mu=0
\eea

\medskip\noindent
Here $J_\mu$ generalises
the global scale symmetry current \cite{F1,F2,F3,F4,F5}.

Let us discuss  the role of  the dilaton; this
is the (log of the) scalar field $\phi$  used to linearise $\hat R$ in
section~\ref{wgsb}. Actually, since we also have the neutral higgs field present, that has non-minimal
coupling to $\hat R$ in the action, the  Stueckelberg field that is also
playing the role of the actual dilaton is now a radial combination of higgs and $\phi$.
Then  the dilaton (in the absence of the $\hat h$)
is actually  replaced by $\phi^2\ra\phi^2+ \xi^2 \,\xi_h \hat h^2$ ($\xi\ll 1$), see (\ref{47}),
while the angular field combination is the real neutral scalar higgs $\sigma$, eq.(\ref{hh}) \cite{DG1}.
For simplicity, below we still refer to $\phi$ as the dilaton (given the small contribution
of $h$).

One can use $\phi$ instead of $\hat R$
as a regulator in (\ref{Sff}), (\ref{op}), as done in eq.(\ref{weyl}),
to keep the Weyl gauge invariance in $d$ dimensions.
As a result of this,  under  the one-loop logarithms in (\ref{Wr}) one can then
replace $\vert\hat R\vert\ra \phi^2$. Actually, this also follows directly by
using the (classical) equation of motion $\phi^2=-\hat R$ ($\hat R<0$), under the logarithm terms
since the  difference of doing so is a higher order quantum correction
(from loop log-like corrections to this equation).

What happens after  breaking of Weyl gauge symmetry,
when $\w_\mu$ becomes massive after `eating' $\ln\phi$
as in Section~\ref{wgsb}? Then $\w_\mu$ decouples (together with $\phi$), and so
$\tilde\Gamma\ra\Gamma$, Weyl connection (geometry)  becomes
Riemannian and $\hat R$, $\hat R_{\mu\nu}$ etc
are replaced by their Riemannian counterparts and $\hat \Box\ra \Box$, with Riemannian
$\Box\!=\!\nabla_\mu\nabla^\mu$. When $\phi$ acquires a vev, ignoring the small
(decoupled)  fluctuations of $\phi$ about $\langle\phi\rangle$, then
eq.(\ref{Wr}) becomes  
\medskip
\bea\label{aaa}
W_r &= &
\int d^4 x \,\sqrt{g}\,\, \Big\{\,
\frac{a_1}{2} R \ln\frac{\Box}{\langle\phi^2\rangle}\, R
+  \frac{b_1}{2}  F_{\mu\nu}  \ln\frac{\Box}{\langle\phi^2\rangle}
\, F^{\mu\nu}+
\frac{c_1}{2} \, C_{\mu\nu\rho\sigma} \,\ln\frac{\Box}{\langle\phi^2\rangle} C^{\mu\nu\rho\sigma}
\nonumber\\[4pt]
&+&
\frac{d_1}{2}
\Big[\, R_{\mu\nu\rho\sigma}\,  \ln{\Box}\, R^{\rho\sigma\mu\nu}
-4  R_{\mu\nu} \,\ln{\Box}\, R^{\nu\mu} + R \,\ln{\Box}\, R\,
\Big] \Big\}+\cdots
\eea

\medskip\noindent
where the dots stand for the tree-level part of the action (terms multiplied by
$a_0$, $b_0$, $c_0$, $d_0$).

In the last line $\ln\langle\phi^2\rangle$ has cancelled out since it multiplied a total derivative.
The above expression of $W_r$  is exactly the Weyl anomaly in Riemannian geometry,
see  \cite{Donoghue} (eq.395). Obviously, the numerical coefficients are the same
in Weyl and Riemannian cases for the same matter content in the loops.

Given the spontaneous breaking of the Weyl gauge symmetry,
it is not too surprising to note  the similarity of   result (\ref{Wr})
to that in \cite{Donoghue} for gravity in Riemannian space-time
with the following natural substitutions in the later, to enforce a
``covariantisation'' with respect to the Weyl gauge symmetry:
$R\ra \hat R$, $R_{\mu\nu}\ra\hat R_{\mu\nu}$, $\Box\ra \hat \Box$, etc.
It seems that a suitable procedure of  ``covariantisation'' of a Riemannian
action and Riemannian operators can lead to a viable implementation of the
gauged dilatations symmetry, as it happens in usual gauge theories of an internal
symmetry. This ends our comparative analysis of Weyl anomaly in theories based
on Riemann vs Weyl conformal geometry.

\subsection{Renormalizability and other quantum operators}\label{ren}

Let us comment on the  renormalizability of  actions (\ref{inA}), (\ref{SMW}), see also (\ref{Sf}).
This  is a gauge theory, so it is quadratic in curvatures terms present.
Eq.(\ref{inA}) is the most general action one can write  perturbatively;
it  coincides with that obtained by
constructing a  gauge theory of the Weyl group in the  modern approach to
gauge theories using the tangent space formulation uplifted to space-time
using the vielbein \cite{CDA}.

In a perturbative approach, in the symmetric phase of this action
there is no higher dimensional counterterm that one could
write,  since no mass scale is allowed in the theory to suppress it,
otherwise the Weyl gauge symmetry would be broken explicitly.
By power counting arguments, the Weyl general action without matter, eq.(\ref{inA}),
is a renormalizable gauge theory. This remains true
if matter is present, eq.(\ref{SMW}.

Regarding the  broken phase  of the action,
the theory is in the so-called  {\it  unitary (physical) gauge}  of the gauged dilatations
  symmetry, in which the physical mass terms are manifest. Just like
the unitary gauge of  SM $U(1)_Y$ of hypercharge, this  is a non-renormalizable gauge.
Hence this  is not a suitable gauge to discuss  renormalizability  - one must use the
symmetric phase to calculate quantum corrections, with a Weyl gauge invariant
regularisation, eq.(\ref{Sf}).

In the Weyl gauge invariant phase of the theory,
counterterms like $\sqrt{g}\,(\hat C_{\mu\nu\rho\sigma}^2)^2/\hat R^2$ are possible,
since the symmetry allows them, although they seem to be non-perturbative. 
How are they generated?  To compute quantum corrections in the symmetric
phase a regularisation that respects this symmetry is needed. We saw 
that such regularisation exists and can be used for loop calculations
to preserve  Weyl gauge symmetry at quantum level.

A lot can be learnt about this from  scale invariant theories in flat spacetime.
In global scale invariant theories in flat space-time it is known that a (global)
scale invariant regularisation is possible, with the aid of a dilaton ($\phi$) that replaces the
DR subtraction scale $\mu$
\cite{Englert,Misha1,Misha2,Misha2s,Misha3,Tamarit,dg1,dg2,dg3,dg7,3loop}
with close  similarities to our analysis above for anomalies.
The quantum corrected action remains  scale invariant and $\mu$
is generated when $\phi$ acquires a vev and the quantum scale symmetry is
broken spontaneously. Then apparently non-perturbative,
non-polynomial higher dimensional counterterms, suppressed by
powers of  $\phi$ are generated, e.g. $(H^\dagger H)^3/\phi^2$, $(H^\dagger H)^4/\phi^4$, etc
\cite{dg1,dg2,dg3}.
These operators respect the scale symmetry of the
action and the Callan-Symanzik equations are verified up to three-loop order \cite{dg7,dg2,dg3}.

A  similar result  may  apply in a theory in curved space-time, as here.
Since the Stueckelberg field $\phi$  was
eaten to all orders by $\w_\mu$, terms suppressed by powers of $\phi$ like
$\hat R^3/\phi^2$, $\hat R^4/\phi^4$, $(\hat C_{\mu\nu\rho\sigma}^2)^2/\phi^4$, etc,
cannot be written, it seems. But with $\phi^2$ replaced onshell by $\vert\hat  R\vert$
(via classical equation $\phi^2=\vert \hat R\vert$), the  situation changes; moreover,
in the light of our discussion with $\hat R^\epsilon$ acting as an equivalent
regulator (eq.(\ref{Sf})), one expects that operators
suppressed by powers of $\hat R$, like
$(\hat C_{\mu\nu\rho\sigma}^2)^2/\hat R^2$ etc, be  generated as counterterms at some loop order; they
respect the Weyl gauge symmetry of the action. Hence, although such non-polynomial operators seem
to be non-perturbative, they may still be captured by a  (quantum) perturbative approach in
a Weyl gauge invariant regularisation, which is interesting.
The impact of  such (apparently non-perturbative) non-polynomial
operators on the renormalizability of the theory was not studied.

Finally, consider the case of  the term
$C_{\mu\nu\rho\sigma}^2$ in (\ref{inA}),  (\ref{EP}), which  is somewhat special\footnote{This
operator is independent of $\omega_\mu$ and has the same form in the symmetric and broken
phases.}. This operator has an additional symmetry (special conformal symmetry),
beyond the Weyl gauge symmetry of the other operators in (\ref{inA}).
Suppose we do not include it in the classical action. A reason for this would be to 
avoid the presence of the associated massive spin-2 state
of mass $\eta \, M_p$ ($\eta\sim 1$) that is generated by $C_{\mu\nu\rho\sigma}^2$
in the presence of the Einstein term \cite{LAG}, in the broken phase eq.(\ref{EP}).
But at quantum level this
higher derivative operator may still be generated as counterterm, like the
$(\hat C_{\mu\nu\rho\sigma}^2)^2/\hat R^2$ term, etc mentioned above.

At  higher orders,  a series of such Weyl gauge invariant
higher derivative operators (suppressed by powers of $\hat R$) may be generated,
of which $C_{\mu\nu\rho\sigma}^2$ is just the first term.
Given the structure of these operators, finding  the
exact  quantum spectrum can become a  non-perturbative issue.
It is possible that the overall sum of this series may actually be ghost-free;
an interesting possibility is that  this series of quantum operators is actually
 re-summed into  the general Weyl-DBI action of Weyl conformal geometry (discussed in
 the next section), which includes at least some of these quantum operators in its expansion.

 Otherwise, in the absence of such re-summation,
the quantum  theory is Weyl gauge invariant at every loop order, but
may have massive states of negative norm (ghosts) - an artefact of
truncating this series (of higher derivative operators)
to that order, be it tree-level or higher loop order.
Such massive states violate unitarity at high scale
(close to their mass) but decouple at low scales  where unitarity is  restored, and are not
produced as final states \cite{HH}. (Renormalizability does not help 
on this: such states also exist in renormalizable quadratic
Riemannian-based gravity \cite{KS} that does not have Weyl gauge symmetry).

Finally, it is often thought that in  theories with higher derivative operators
the UV behaviour is actually improved by such operators. The Weyl gauge symmetry
present here may also help. But the outcome depends on the details
of the  Wick rotation to Euclidean space which can be non-trivial in this case
(the UV behaviour  is not always improved, see e.g. \cite{ghosts}).
This issue deserves further study.

 \subsection{Weyl quadratic gravity from  Weyl gauge invariant DBI action}

It is natural to ask whether there exists a Weyl gauge invariant action more general than
Weyl quadratic gravity,  eqs.(\ref{Sf0}), (\ref{Sf}), (\ref{inA}).
Surprisingly, such an action exists and is the analogue of the
 Dirac-Born-Infeld action \cite{BI,D,Sorokin} for  Weyl conformal geometry, as it was shown
in \cite{DBI}. Here we review briefly this result.
 To find this action, note from (\ref{tra}) that
$\hat R g_{\mu\nu}$, $\hat R_{\mu\nu}$, $\hat F_{\mu\nu}$ are invariants of (\ref{WGS}) - hence we can
 construct a Weyl - DBI action (hereafter WDBI) in  $d$ dimensions as follows \cite{DBI}
\bea\label{dbi}
S_{\rm WDBI}=\int d^d\sigma\Big\{-
  \det\,[s_0 \,\hat R\, g_{\mu\nu}+s_1 \,\hat R_{\mu\nu}+ s_2 \,\hat F_{\mu\nu}]\Big\}^{\frac12}
\eea

\medskip\noindent
where $s_0, s_1, s_2$ are dimensionless constants. Note that no UV regulator or DR
subtraction scale ($\mu$) is required here, even though this action is actually defined
in $d$ dimensions and is Weyl gauge invariant.
The  analytical continuation of this action when going  from $d=4$ to $d$
dimensions  is trivial (simply replace $d\!=\!4$ by arbitrary  $d$).
Unlike the usual DBI actions, all terms here involve derivatives (including
the first term that multiplies $g_{\mu\nu}$). Further
\medskip
\bea
S_{\rm WDBI}=
\int d^d\sigma
\sqrt{g}\,\, (s_0\,\hat R)^{\frac{d}{2}}\,\,
\Big\{  \det \,\big[\delta_{\,\,\,\nu}^\lambda+X^\lambda_{\,\,\,\nu}\,\big]\Big\}^{\frac12},
\quad
X^\lambda_{\,\,\,\,\nu}= \frac{1}{s_0\, \hat R}\, g^{\lambda\rho} 
\big[s_1\,\hat R_{\rho\nu}+  s_2 \,\hat F_{\rho\nu}\,\big].
\eea

\medskip\noindent
Under the assumption $\vert s_{1,2}/s_0\vert\ll 1$ we can expand the action and use that
\medskip
\bea
\big[ \det \big(1+X\big)\big]^{\frac12}\!\!
 &=&\!\!
1+\frac12\, \tr X+\frac14\,
\Big[\frac12 (\tr X)^2-\tr X^2\Big] +
\nonumber\\[4pt]
 &+&\!\!\!\!
 \Big[\frac{1}{48} (\tr X)^3-\frac{1}{8} \tr X \,\tr X^2
  +\frac16\,\tr X^3\Big]
  +\cO(X^4).
\eea

\medskip\noindent
We have
\bea\label{traces}
\tr X&=& \frac{s_1}{s_0},\qquad
\tr X^2 = \frac{1}{s_0^2\,\hat R^2} \Big[
 s_1^2\, \hat R_{\mu\nu}\hat R^{\nu\mu}
+s_2 \Big( s_2 + s_1\,\alpha\,q\,\frac{d-2}{2}\Big) \hat F_{\mu\nu} \hat F^{\nu\mu}\Big]
\\[5pt]
\tr X^3 &= & \frac{1}{\hat R^3} \, \Big(\frac{s_1}{s_0}\Big)^3 \hat R^{\mu\sigma} \hat R_{\sigma\rho}\,
\hat R_{\nu\mu} \, g^{\nu\rho} + \cdots
\eea

\medskip\noindent
With (\ref{id2}) we replace the  $\hat R_{\mu\nu} \, \hat R^{\nu\mu}$ contribution
 in terms of $\hat C_{\mu\nu\rho\sigma}^2$ and the Euler term $\hat G$.
Bringing everything together
\medskip
\bea\label{S3}
S_{{\rm WDBI}}=\int d^d\sigma \sqrt{g}\,\,  (\hat R^2)^{\frac{d}{4}-1}\, \Big[\,
  \frac{1}{4! \,\xi^2}  \,\hat R^2-\frac{1}{\eta^2}
  \hat C_{\mu\nu\rho\sigma}\hat C^{\mu\nu\rho\sigma}
  -\frac{1}{\zeta}
  \hat F_{\mu\nu} \, \hat F^{\mu\nu}+\frac{1}{\eta^2}\, \hat G +\cO(X^3)\Big]
\eea
where
\medskip
\bea\label{couplings}
\frac{1}{4!\, \xi^2}&=&
\Big[\, s_0^2+ \frac12 \,s_0\,s_1+ \,s_1^2\,\frac{d-2}{16 \,(d-1)}\,\Big]\,s_0^{\frac{d}{2}-2}\,;
\nonumber\\[6pt]
\frac{1}{\eta^2}&= &
\frac{1}{16} \frac{d-2}{(d-3)}\,s_1^2 \,s_0^{\frac{d}{2}-2},\quad
\frac{1}{\zeta}=-\frac14 \,s_2\,\Big[ s_2+\frac12 s_1 \alpha \, q\, (d-2)\Big] s_0^{d/2-2}.
\qquad
\eea

\medskip
Therefore, the WDBI action  in the leading order recovered the Weyl quadratic gravity action
in $d$ dimensions eq.(\ref{Sf}) that is already suitably regularised
such as to preserve the Weyl gauge symmetry.
While action (\ref{Sf}) was introduced  on symmetry grounds as a geometric
analytical continuation to $d=4-2\epsilon$ to maintain Weyl gauge invariance,
here  the WDBI action in $d$ dimensions automatically  generates such Weyl gauge invariant
regularisation. This is an interesting result that
supports the  regularisation used for Weyl anomaly (\ref{op}) based on 
the analytical continuation in eq.(\ref{Sf}), (\ref{Sff}), with scalar curvature
$\hat R$ as regulator.

Let us consider the limit $d=4$. Action (\ref{S3}) becomes
\bea\label{ss4}
S_{\rm WDBI,4}&=&\int d^4\sigma
\Big\{-
  \det\,[s_0 \,\hat R\, g_{\mu\nu}+s_1 \,\hat R_{\mu\nu}+ s_2 \,\hat F_{\mu\nu}]\,\Big\}^\frac12
\\
&=&
\int d^4\sigma \sqrt{g}\, \Big[\,
  \frac{1}{4! \,\xi^2}  \,\hat R^2-\frac{1}{\eta^2}
  \hat C_{\mu\nu\rho\sigma}\hat C^{\mu\nu\rho\sigma}
  -\frac{1}{\zeta}
  \hat F_{\mu\nu} \, \hat F^{\mu\nu}+\frac{1}{\eta^2}\, \hat G +\cO(X^3)\Big],
\label{S4}\eea
with
\bea\label{couplings4}
\frac{1}{4!\, \xi^2}&=&
s_0^2+ \frac12 \,s_0\,s_1+ \,\frac{1}{24}\,s_1^2,
\qquad
\frac{1}{\eta^2}=
\frac18\,s_1^2,
\qquad
\frac{1}{\zeta}=-\frac14 \, s_2 (s_2+s_1 \alpha\,q).
\eea

\medskip
Therefore, the WDBI action eq.(\ref{ss4}) exactly  recovers in (\ref{S4})
the initial Weyl quadratic gravity action  (\ref{inA}) in $d=4$,
under the following assumptions:
convergence constraints of the expansion $\vert s_{1,2}\vert\ll \vert s_0\vert$,
well-defined gauge kinetic term of $\w_\mu$,  $\zeta>0$, and perturbativity
in action (\ref{inA}). These constraints are respected
if we take $\xi\ll \eta<1$ and $s_2$ is chosen to satisfy
$-s_1 \alpha q<s_2<0$ or $0<s_2<-s_1 \alpha q$ for $\zeta>0$.
The condition $\xi\ll \eta<1$ allows  the mass of spin-two state
from the Weyl term $C_{\mu\nu\rho\sigma}^2$ in Weyl quadratic action be close to the Planck scale
$m_\eta=\eta\,M_p\sim M_p$, so that below $M_p$ the Einstein-Proca action
dominates (see eq.(\ref{EP}) and the new paragraph after eq.(\ref{omega-limit})).
Since  Einstein-Hilbert gravity is recovered in the  broken phase of
Weyl quadratic gravity, it is  also obtained in the broken phase of this
WDBI gauge theory, eq.(\ref{ss4}). This is 
a purely geometric result: again no matter fields or Weyl scalar compensators
were added ``by hand''.  In general, the WDBI action can have more
general couplings in (\ref{ss4}) than  Weyl quadratic gravity.

What about $\cO(X^3)$ and higher corrections from the expansion?
these terms are suppressed by Weyl scalar curvature, e.g.
$\hat R^{\mu\sigma} \hat R_{\sigma\rho} \hat R^{\rho}_{\,\,\,\,\mu}/\hat R$ or 
$(\hat C_{\mu\nu\rho\sigma}^2)^2/\hat R^2$, where  $\hat C^2_{\mu\nu\rho\sigma}$ has all
indices contracted by the metric. These terms when multiplied by $\sqrt{g}$ are 
Weyl gauge invariant and are important when the expansion is slowly convergent,
$\hat R$ is small, etc. The WDBI action sums up all such apparently non-perturbative terms.
They remind us about similar non-polynomial operators discussed in Section~\ref{ren} that
can be generated at the quantum level, by a Weyl gauge invariant
regularisation, with $\hat R$ as regulator (replacing the usual DR subtraction
scale $\mu$). These terms have similar structure and symmetry.

Briefly, the Weyl gauge invariant WDBI action sums up at least some of the non-polynomial
(higher derivative) quantum corrections to Weyl quadratic gravity, mentioned in
Section~\ref{ren}, while its leading order 
 is exactly the original Weyl quadratic gravity action! The relation of the WDBI action 
 to the gauge invariant
 quantum corrections of Weyl quadratic gravity deserves further study.

There is another interesting limit of  WDBI action eq.(\ref{ss4}),
when $\w_\mu$ is ``pure gauge'' so its field strength vanishes;
formally this means $s_2=0$  and the action becomes
\medskip
\bea
S_{\rm WDBI, 4}&=&\int d^4\sigma\Big\{-
  \det\,[s_0 \,\hat R\, g_{\mu\nu}+s_1 \,\hat R_{\mu\nu}]\Big\}^{\frac12}
\nonumber\\
&=&
\int d^4\sigma \sqrt{g}\, \Big[
  \frac{1}{24 \xi^2}  \,\hat R^2-\frac{1}{\eta^2}
  \hat C_{\mu\nu\rho\sigma}\hat C^{\mu\nu\rho\sigma}
  +\cO(X^3)\Big].
\eea

\medskip\noindent
This action simplifies  when going to the Riemannian picture  of the broken phase
(Einstein gauge/frame) which is the physical one. We first linearise $\hat R^2$ term
as done earlier in this review, $\hat R^2\ra -2 \phi^2 \hat R-\phi^4$ which
when used in the action leads to an equation of motion of solution $\phi^2=-\hat R$ ($\hat R<0$).
Using (\ref{Ri2}), (\ref{nat}), (\ref{ccc}),  one expresses
$\hat R$ in terms of its Riemannian notation while the Weyl
tensor term does not change when going to the Riemannian picture; one can
then integrate out  algebraically the field $\w_\mu$. The resulting action
in Riemannian notation is exactly action (\ref{lim}) plus  $\cO(X^3)$ corrections!
When $\phi$ acquires a vev, Einstein-Hilbert action is generated
by  $\phi^2  R$, with Planck scale as in (\ref{la}).
Note that here  the dilaton action coupled to gravity
is {\it not} added by hand but is of geometric origin from the $\hat R^2$
term. The action has local  Weyl symmetry
(there is no $\w_\mu$) and arises from the DBI action in the integrable geometry limit,
just like (\ref{lim}) is a particular limit of (\ref{inA}).  

One can also consider a $U(1)$ gauge symmetry in the
WDBI action. This means adding its field strength $\cF_{\mu\nu}$ contribution,
with a given coefficient to the matrix under the determinant of WDBI action.
Weyl gauge symmetry remains present,  for details see \cite{DBI}.
Further, given that a DBI action with a $U(1)$  is,
under some assumption, close to a low energy effective description
of a D-brane action in string theory, one may ask how close such WDBI
action is to a D3-brane action in the background of closed string modes:
$G_{\mu\nu}$, the antisymmetric two-form $B_{\mu\nu}$ and the dilaton $\Phi$
\cite{Tasi}
\bea
S_{D3}=
T_3\int d^{4} \xi e^{-\Phi}\sqrt{-\det (\gamma_{ab}+2\pi\alpha^\prime \cF_{ab}+B_{ab})}
\eea
Here $\gamma_{ab}$ is the pull-back metric defined by $G_{\mu\nu}$ (similar for $B_{ab}$).
The problem is that Weyl gauge symmetry is not a symmetry
in strings - the brane tension breaks it. However, this tension is generated by the dilaton
vev, much like the Planck scale is generated in the broken phase of (\ref{inA}) 
(in the leading order of WDBI action). An analogue of
$B_{\mu\nu}$  on conformal geometry side could  be $\hat F_{\mu\nu}$ of $\w_\mu$.
Note however, that while in the D3-brane action there exists a three-form $H=d B$ that
plays the role of torsion tensor, on the conformal geometry side torsion is only vectorial
\cite{CDA}. One should then consider an extension of Weyl conformal geometry in the
presence of a totally antisymmetric torsion tensor \cite{CDA2}.
It may be worth exploring a possible connection of the WDBI action to the D3-brane action.

\subsection{Other implications}

There are other important physical implications of Weyl geometry that we did not address.

One of these  may relate to the interesting `conformal cyclic cosmology' (CCC) model
\cite{Penrose}, based on conformal geometry. However in this model there is no Weyl
gauge boson associated.  It seems to us that a  natural framework for
such model would be general Weyl conformal geometry,
with dynamical $\w_\mu$ present, as discussed above. That would allow a  full geometric
interpretation of the associated gravity theory, as we discussed, with
an underlying gauge symmetry. Transitions from one aeon to another, by
conformal transformation, would then be extended to include matching
conditions for  $\w_\mu$, to benefit from the power of the gauge
symmetry of Weyl conformal geometry. 

Another interesting issue is that of a candidate for dark matter -
the Weyl gauge field $\w_\mu$ can be such dark matter candidate \cite{Huang,Tang}.
In the light of the origin of $\w_\mu$ in Weyl geometry, this would give a geometric
solution to the dark matter problem, while from the Riemannian perspective,
$\omega_\mu$ is just another dof, giving a WIMP solution,
and thus reconciling these two views. Interestingly, in  \cite{Harko} it was
found that a simple solution of  Weyl quadratic gravity
(in the presence of a radial component  of $\w_\mu$) can successfully account
for the large variety of the rotation curves of the SPARC sample considered there,
for 175 galaxies.

Finally, a possible  implication of Weyl conformal geometry-based theories
relates to Big-Bang and black-hole singularities: in Einstein-Hilbert gravity,
the existence of such singularities can indicate the geodesic incompleteness
of the theory. Such  singularities are often  thought to be solved 
by a theory of  quantum gravity. If so, in  a  Weyl geometry-based
quantum gravity theory, one could expect that such singularities
be absent. This is actually consistent with the expectation that Weyl conformal geometry
is geodesic complete, for a discussion see \cite{Ehlers,Ohanian}.

\section{Conclusions}

Weyl conformal geometry is the underlying geometry of the
Weyl quadratic gravity action which 
is a gauge theory of the Weyl group (of dilatations and Poincar\'e symmetry).
In its Weyl gauge covariant formulation, Weyl geometry is simply
 a ``covariantised'' version of Riemannian geometry (of its connection, etc)
 with respect to the gauged dilatations symmetry, something
 overlooked in the literature.
In this  covariant formulation, which is the only physical one from a gauge theory
perspective, Weyl geometry is actually {\it metric},
$\hat\nabla_\mu g_{\alpha\beta}=0$. This is important since one can then
do classical and quantum calculations directly in this geometry,
without going to a (metric) Riemannian formulation, as usually done. This
theory is  a  (quadratic) vector-tensor gauge theory of gravity,
with a conserved Weyl gauge current.

In the limit the Weyl gauge current is trivial (vanishes),  Weyl geometry
becomes  Riemannian and the action  becomes that of  conformal gravity
(given by the Weyl-tensor-squared) plus that of a conformally coupled scalar field.
While the conformal gravity action  can itself be obtained from gauging
the full conformal group, it is not a true gauge theory since no dynamical (physical)
gauge bosons of dilatations and of special conformal symmetry are present
in such action.
This explains  why this action is recovered for a vanishing Weyl  current, as
 discussed.
As a result, Weyl quadratic gravity is the only true gauge theory of a space-time symmetry
beyond Poincar\'e symmetry, with a  geometric interpretation (for the origin
of its degrees of freedom and mass scales). 
 
The  gauge symmetry of Weyl quadratic gravity is
broken \`a la Stueckelberg, the  Weyl gauge field $\w_\mu$ becomes massive (by eating
$\ln\phi$, where $\ln\phi$  is the would-be-Goldstone propagated by $\hat R^2$); then
$\w_\mu$ decouples,
Weyl geometry (connection) becomes Riemannian (Levi-Civita connection), respectively, 
and we recover Einstein-Hilbert action with  $\Lambda>0$.
All mass scales of the theory $M_p$, $\Lambda$, $m_\w$ are  generated by the vev of
$\phi$ which means  they all have a {\it geometric}
origin. This is an elegant explanation for the origin of mass, with no ``moduli''
fields added ``by hand'' to this purpose. Also, unlike other theories,
here there is no need to stabilize the vev of $\phi$, since it is eaten by $\omega_\mu$,
thus elegantly  avoiding this problem. 
The  field $\phi$  also plays the role of inflaton (scalaron) in an inflation scenario,
showing the multiple roles of this field. And with Riemannian geometry (Levi-Civita
connection) recovered in the broken phase (after massive $\omega_\mu$ decouples), we see
that Weyl geometry is more fundamental for physics.

The above results  easily avoid the century-old (wrong) argument against the physical
relevance of Weyl conformal geometry, that invoked the change of
norm of vectors under their parallel transport,
due to its apparent non-metricity i.e. non-zero $\tilde\nabla_\mu g_{\alpha\beta}\not=0$.
In particular this argument implied that the atomic lines spacing would be path
history dependent (second clock effect). In fact, this argument does not apply
to the Weyl gauge theory, when: 1) one  implements  manifest {\it Weyl gauge  covariance}
of the parallel transport of vectors, for this to actually be physically relevant,
and 2): one shows  that  the Weyl gauge boson is actually massive.
As a result,  the symmetric phase has no second-clock effect (there
is no mass scale to generate a clock rate in the first instance) and the theory
is Weyl covariantly metric ($\hat \nabla_\mu g_{\alpha\beta}=0$) and the norm of vectors
is then  invariant; while in the spontaneously broken phase any such effect that one might 
insist to exist is suppressed by the (large) mass of the Weyl gauge field.
These arguments avoid the century-old criticism against a physical relevance of
Weyl conformal geometry and open new directions of research.

What happens when one adds matter to Weyl geometry? 
 The above picture remains valid, with some corrections.
The SM with vanishing higgs mass {\it parameter} is scale invariant, hence it
can have a truly minimal, natural embedding in Weyl geometry, with no
new degrees of freedom  beyond those of Weyl geometry and SM.
As a result, only
the Higgs sector acquires a tree-level coupling to the Weyl gauge boson,
through the Weyl gauge covariant derivative of the kinetic term,
with possible phenomenological relevance, yet to be studied.
Weyl gauge symmetry may actually play a role
in providing a technically natural solution to the hierarchy problem.
Regarding the SM gauge bosons and fermions, they do not have tree-level couplings to $\w_\mu$,
unless a small gauge kinetic mixing $\w_\mu$-hypercharge is present.

The SM in Weyl geometry accommodates  a natural inflation scenario.
Inflation is of Starobinsky-Higgs type, since the model is essentially a gauged version
of Starobinsky inflation, driven essentially by the ``scalaron'' (dilaton $\phi$).
The predicted value for the tensor-to-scalar
ratio $0.000227\leq r\leq 0.00303$ ($95\%$ CL)  is bounded
from above by that of the Starobinsky model which is saturated for a vanishing  Higgs
coupling to Weyl geometry. Assuming the inflationary scenario,
this is a clear, narrow prediction of SM in  Weyl geometry that can be tested by future
experiments.

With dilatations as a gauge symmetry, it is important that this symmetry be anomaly-free, so
that Weyl quadratic gravity associated to Weyl geometry is a consistent (quantum) gauge theory
(of this symmetry). Remarkably, in Weyl geometry there is no Weyl anomaly.
The main reason is that the Weyl term  $C_{\mu\nu\rho\sigma}^2$
{\it and} the Euler term $\hat G$ in the action are both Weyl gauge covariant.
This was true anyway for  $C_{\mu\nu\rho\sigma}^2$ in the Riemannian picture,
but that $\hat G$ is Weyl gauge covariant is the new, specific feature 
of  Weyl geometry.  Due to this,  a  natural Weyl gauge
invariant  {\it geometric} regularisation is then possible for the action;
 there is {\it no UV regulator/subtraction scale} $\mu$ -
its role is played by the Weyl-covariant scalar curvature  $\hat R\not=0$
of Weyl geometry. In this way, Weyl gauge invariance is manifestly
maintained in $d=4-2\epsilon$ dimensions and in the one-loop action. The result is that
the anomaly associated usually to both the Weyl and Euler terms 
is absent, also due to the additional presence of a dynamical
degree of freedom (dilaton/would-be-Goldstone of dilatations).
The anomaly is recovered in the broken phase
after the dilaton decouples (with massive $\w_\mu$).

We also discussed an elegant  generalisation of  Weyl quadratic gravity action;
we presented an analogue of the Dirac-Born-Infeld action,  for Weyl  conformal geometry,
called Weyl-DBI action (WDBI) that is  Weyl gauge invariant in arbitrary $d$ dimensions,
with dimensionless couplings. The WDBI action
is  special: it has  no need for a UV regularisation and
UV subtraction scale - the Weyl scalar curvature plays
the  role of UV regulator and maintains  Weyl gauge symmetry in $d$ dimensions.
There is no need for an analytical continuation from $d=4$ to $d$ dimensions
in this theory (the continuation  is trivial, simply replace $d=4$ by arbitrary
$d$ in the action). The original (regularised)  Weyl quadratic gravity action\footnote{with a
Weyl gauge invariant regularisation in $d=4-2\epsilon$.} was shown to be the leading order
of an expansion\footnote{in powers of ratios of the dimensionless couplings.}
of  this WDBI action in $d$ dimensions; the WDBI contains in addition
a series of higher derivative operators suppressed by powers of $\hat R$,
like $(C_{\mu\nu\rho\sigma}^2)^2/\hat R^2$, etc. Such apparently non-perturbative
 operators can also be  generated by perturbative quantum calculations
 in Weyl quadratic gravity (in a Weyl gauge invariant regularisation),
 with $C_{\mu\nu\rho\sigma}^2$ as the first term in this series.
The WDBI action sums up at least  some of these quantum operators,
which is interesting. 

Briefly, Weyl conformal geometry provides a good candidate for a
(quantum) gauge theory of gravity, represented by Weyl quadratic gravity
and leads to a unified description, by the gauge principle,
of Einstein-Hilbert gravity and SM interactions.

\bigskip
\begin{center}
  ---------------------------------------
\end{center}

\vspace{0.5cm}
\noindent
{\bf Acknowledgements:\,\,\,}

\noindent
The author thanks Cezar Condeescu and Andrei Micu (IFIN Bucharest) for
interesting discussions on Weyl conformal geometry.



\begin{thebibliography}{100}{
\bibitem{Donoghue}
J.~F.~Donoghue, M.~M.~Ivanov and A.~Shkerin,
``EPFL Lectures on General Relativity as a Quantum Field Theory,''
[arXiv:1702.00319 [hep-th]].

  \bibitem{Noether}
    E. Noether, “Invariant Variation Problems,” Gott. Nachr. 1918 (1918) 235 [Transp.
      Theory Statist. Phys. 1 (1971) 186] doi:10.1080/00411457108231446 [physics/0503066]. Gott.
Nachr. 1918 (1918) 235

  \bibitem{Self}
   Affine transformation: see e.g.  Wolfram Mathworld
   ``The makers of Mathematica and Wolfram Alpha''
   https://mathworld.wolfram.com/AffineTransformation.html

 \bibitem{Koch}
 Koch curve, see for example:   Wolfram Mathworld
``The makers of Mathematica and Wolfram Alpha''
http://mathworld.wolfram.com/KochSnowflake.html;

 \bibitem{si}
   Y.Khaluf, E. Ferrante, P. Simoens, C, Huepe, “Scale invariance in natural and artificial
collective systems: a review”, Journal of Royal Society “Interface” 14: 20170662


\bibitem{astro} ``The Millennium Simulation Project'',
  Max Planck Institut fur Astrophysik, see 
  https://wwwmpa.mpa-garching.mpg.de/galform/virgo/millennium/
   

\bibitem{Mandelbrot}
   B. B. Mandelbrot, “The Fractal Geometry Of Nature”. San Francisco 1982.

\bibitem{N1}
  L. Nottale, ``Scale relativity and fractal space-time - a new approach to
  unifying relativity and quantum mechanics'',
Imperial College Press, ISBN 978-1-84816-650-9, 2011.

\bibitem{t1}
  G. ’t Hooft,
``Local conformal symmetry: The missing symmetry component for space and time,''
International Journal of Modern Physics D \textbf{24} (2015) no.12, 1543001
doi:10.1142/S0218271815430014, arXiv:1410.6675 [gr-qc].

\bibitem{t4}
G.~'t Hooft,
``Local conformal symmetry in black holes, standard model, and quantum gravity,''
International Journal of Modern Physics D \textbf{26} (2016) no.03, 1730006
doi:10.1142/S0218271817300063


\bibitem{t2}
  G. ’t Hooft, “A class of elementary particle models without any adjustable real parameters,” Found. Phys. 41 (2011) 1829 [arXiv:1104.4543 [gr-qc]].

\bibitem{t3}
  I. Bars, P. Steinhardt and N. Turok, “Local Conformal Symmetry in Physics and Cosmology,”
  Phys. Rev. D 89 (2014) no.4, 043515 [arXiv:1307.1848 [hep-th]].

\bibitem{Misha1}
  M. Shaposhnikov and D. Zenhausern, “Quantum scale invariance, cosmological constant
and hierarchy problem,” Phys. Lett. B 671 (2009) 162 [arXiv:0809.3406 [hep-th]].

\bibitem{Englert}
F.~Englert, C.~Truffin and R.~Gastmans,
``Conformal Invariance in Quantum Gravity,''
Nucl. Phys. B \textbf{117} (1976), 407-432
doi:10.1016/0550-3213(76)90406-5

\bibitem{Kaku}
M.~Kaku, P.~K.~Townsend and P.~van Nieuwenhuizen,
``Gauge Theory of the Conformal and Superconformal Group,''
Phys. Lett. B \textbf{69} (1977), 304-308
doi:10.1016/0370-2693(77)90552-4

\bibitem{mannheim}
P.~D.~Mannheim and D.~Kazanas,
``Exact Vacuum Solution to Conformal Weyl Gravity and Galactic Rotation Curves,''
Astrophys. J. \textbf{342} (1989), 635-638
doi:10.1086/167623;
 D.~Kazanas and P.~D.~Mannheim,
``General Structure of the Gravitational Equations of Motion in Conformal Weyl Gravity,''
Astrophys. J. Suppl. \textbf{76} (1991), 431-453
doi:10.1086/191573

\bibitem{vanProeyen}
D.~Z.~Freedman and A.~Van Proeyen,
``Supergravity,''
Cambridge Univ. Press, 2012,
ISBN 978-1-139-36806-3, 978-0-521-19401-3
doi:10.1017/CBO9781139026833

\bibitem{Tait}
J.~M.~Charap and W.~Tait,
``A gauge theory of the Weyl group,''
Proc. Roy. Soc. Lond. A \textbf{340} (1974), 249-262
doi:10.1098/rspa.1974.0151

\bibitem{Weyl1}
Hermann Weyl, Gravitation und elektrizit\"at, Sitzungsberichte der
K\"oniglich Preussischen Akademie der Wissenschaften zu Berlin (1918), pp.465.
This work includes     Einstein's report appended, stating that
the  atomic spectral lines spacing changes under parallel transport,
thus leading to the dependence of this distance on the path
history of each atom due to the non-metricity of the underlying geometry
($\tilde\nabla_\mu g_{\alpha\beta}\not=0$),
in contrast with experience (so-called second clock effect).
(An English version by D. H. Delphenich is currently available at:
http://www.neo-classical-physics.info/spacetime-structure.html)


\bibitem{Weyl2}
Hermann Weyl ``Eine neue Erweiterung der Relativitätstheorie'' 
(``A new extension of the theory of relativity''), Ann. Phys. (Leipzig) (4) 59 (1919), 101-133.
(an English version by D.H. Delphenich is currently available at this link:
http://www.neo-classical-physics.info/spacetime-structure.html)


\bibitem{Weyl3}
Hermann Weyl ``Raum, Zeit, Materie'', vierte erweiterte Auflage. Julius Springer, Berlin 1921
``Space-time-matter'', translated from German by Henry L. Brose, 1922, Methuen \& Co Ltd, London,
www.gutenberg.org/ebooks/43006 (Project Gutenberg License).


\bibitem{CDA}
C.~Condeescu, D.~M.~Ghilencea and A.~Micu,
``Weyl quadratic gravity as a gauge theory and non-metricity vs torsion duality,''
Eur. Phys. J. C \textbf{84} (2024) no.3, 292
doi:10.1140/epjc/s10052-024-12644-6
[arXiv:2312.13384 [hep-th]].

\bibitem{DG1}
D.~M.~Ghilencea,
``Weyl conformal geometry vs Weyl anomaly,''
JHEP \textbf{10} (2023), 113
doi:10.1007/JHEP10(2023)113
[arXiv:2309.11372 [hep-th]];

\bibitem{Scholz}
E.~Scholz,
``The unexpected resurgence of Weyl geometry in the late 20-th century physics,''
Einstein Stud. \textbf{14} (2018), 261-360
doi:10.1007/978-1-4939-7708-6\_11
[arXiv:1703.03187 [math.HO]].


\bibitem{Lasenby}
M.~P.~Hobson and A.~N.~Lasenby,
``Weyl gauge theories of gravity do not predict a second clock effect,''
Phys. Rev. D \textbf{102} (2020) no.8, 084040
doi:10.1103/PhysRevD.102.084040
[arXiv:2009.06407 [gr-qc]].

\bibitem{non-metricity}
D.~M.~Ghilencea,
``Non-metric geometry as the origin of mass in gauge theories of scale invariance,''
Eur. Phys. J. C \textbf{83} (2023) no.2, 176
doi:10.1140/epjc/s10052-023-11237-z
[arXiv:2203.05381 [hep-th]].

\bibitem{D2}
P.~A.~M.~Dirac,
``Long range forces and broken symmetries,''
Proc. Roy. Soc. Lond. A \textbf{333} (1973), 403-418
doi:10.1098/rspa.1973.0070

\bibitem{Ghilen0}
  D.~M.~Ghilencea,
  ``Spontaneous breaking of Weyl quadratic gravity to Einstein action and Higgs potential,''
  JHEP {\bf 1903} (2019) 049
  [arXiv:1812.08613 [hep-th]].
  D.~M.~Ghilencea,
``Stueckelberg breaking of Weyl conformal geometry and applications to gravity,''
  Phys.\ Rev.\ D {\bf 101} (2020) 4,  045010
  [arXiv:1904.06596 [hep-th]].
  For a brief  review see Section 2.1 in \cite{SMW}.

\bibitem{SMW}
D.~M.~Ghilencea,
``Standard Model in Weyl conformal geometry,''
Eur. Phys. J. C \textbf{82} (2022) no.1, 23
 doi:10.1140/epjc/s10052-021-09887-y
[arXiv:2104.15118 [hep-ph]];


\bibitem{Duff}
M.~J.~Duff,
``Twenty years of the Weyl anomaly,''
Class. Quant. Grav. \textbf{11} (1994), 1387-1404
doi:10.1088/0264-9381/11/6/004
[arXiv:hep-th/9308075 [hep-th]].
M.~J.~Duff,
``Observations on Conformal Anomalies,''
Nucl. Phys. B \textbf{125} (1977), 334-348
doi:10.1016/0550-3213(77)90410-2.

\bibitem{Duff2}
D.~M.~Capper, M.~J.~Duff and L.~Halpern,
``Photon corrections to the graviton propagator,''
Phys. Rev. D \textbf{10} (1974), 461-467
doi:10.1103/PhysRevD.10.461

\bibitem{Duff3}
D.~M.~Capper and M.~J.~Duff,
``Trace anomalies in dimensional regularization,''
Nuovo Cim. A \textbf{23} (1974), 173-183
doi:10.1007/BF02748300


\bibitem{Deser1976}
S.~Deser, M.~J.~Duff and C.~J.~Isham,
``Nonlocal Conformal Anomalies,''
Nucl. Phys. B \textbf{111} (1976), 45-55
doi:10.1016/0550-3213(76)90480-6

\bibitem{Deser}
S.~Deser and A.~Schwimmer,
``Geometric classification of conformal anomalies in arbitrary dimensions,''
Phys. Lett. B \textbf{309} (1993), 279-284
doi:10.1016/0370-2693(93)90934-A
[arXiv:hep-th/9302047 [hep-th]].


\bibitem{BI}
   M.~Born and L.~Infeld,
  ``Foundations of the new field theory,''
Proc. Roy. Soc. Lond. A \textbf{144} (1934) no.852, 425-451
doi:10.1098/rspa.1934.0059

\bibitem{D}
 P.~A.~M.~Dirac,
``An Extensible model of the electron,''
Proc. Roy. Soc. Lond. A \textbf{268} (1962), 57-67
doi:10.1098/rspa.1962.0124

\bibitem{Sorokin}
For a review, see D.~P.~Sorokin,
``Introductory Notes on Non-linear Electrodynamics and its Applications,''
Fortsch. Phys. \textbf{70} (2022) no.7-8, 2200092
doi:10.1002/prop.202200092
[arXiv:2112.12118 [hep-th]].
  
\bibitem{Gibbons}
For a review in relation to string theory: G.~W.~Gibbons,
``Aspects of Born-Infeld theory and string / M theory,''
AIP Conf. Proc. \textbf{589} (2001) no.1, 324-350
doi:10.1063/1.1419338
[arXiv:hep-th/0106059 [hep-th]].

\bibitem{WI3}
P.~G.~Ferreira, C.~T.~Hill, J.~Noller and G.~G.~Ross,
``Scale-independent $R^2$ inflation,''
Phys. Rev. D \textbf{100} (2019) no.12, 123516
 doi:10.1103/PhysRevD.100.123516
[arXiv:1906.03415 [gr-qc]].

\bibitem{WI1}
D.~M.~Ghilencea,
``Weyl R$^{2}$ inflation with an emergent Planck scale,''
JHEP \textbf{10} (2019), 209
doi:10.1007/JHEP10(2019)209
[arXiv:1906.11572 [gr-qc]].

\bibitem{WI2}
D.~M.~Ghilencea,
``Gauging scale symmetry and inflation: Weyl versus Palatini gravity,''
Eur. Phys. J. C \textbf{81} (2021) no.6, 510
doi:10.1140/epjc/s10052-021-09226-1
[arXiv:2007.14733 [hep-th]].

\bibitem{Starobinsky}
A.~A.~Starobinsky,
``A New Type of Isotropic Cosmological Models Without Singularity,''
Phys. Lett. B \textbf{91} (1980), 99-102
doi:10.1016/0370-2693(80)90670-X

\bibitem{Harko}
M.~Cr\u{a}ciun and T.~Harko,
``Testing Weyl geometric gravity with the SPARC galactic rotation curves database,''
[arXiv:2311.16893 [gr-qc]];
P.~Burikham, T.~Harko, K.~Pimsamarn and S.~Shahidi,
``Dark matter as a Weyl geometric effect,''
Phys. Rev. D \textbf{107} (2023) no.6, 064008
doi:10.1103/PhysRevD.107.064008
[arXiv:2302.08289 [gr-qc]];

\bibitem{Harko2}
J.~Z.~Yang, S.~Shahidi and T.~Harko,
``Black hole solutions in the quadratic Weyl conformal geometric theory of gravity,''
Eur. Phys. J. C \textbf{82} (2022) no.12, 1171
doi:10.1140/epjc/s10052-022-11131-0
[arXiv:2212.05542 [gr-qc]].


\bibitem{Ehlers}
 J. Ehlers, F. A. E. Pirani and A. Schild, ``Republication of:
 The geometry of free fall and light propagation'', Gen Relativ Gravit
 (2012) 44:1587–1609, doi: 10.1007/s10714-012-1353-4.
 Original paper: J. Ehlers, F. A. E. Pirani and A. Schild
 in: General Relativity, papers in honour of J. L. Synge.
 Edited by L. O’Raifeartaigh. Oxford, Clarendon Press 1972, pp. 63–84.
 
\bibitem{Ohanian}
H.~C.~Ohanian,
``Weyl gauge-vector and complex dilaton scalar for conformal symmetry and its breaking,''
Gen. Rel. Grav. \textbf{48} (2016) no.3, 25
doi:10.1007/s10714-016-2023-8
[arXiv:1502.00020 [gr-qc]].


\bibitem{Smolin}
 L.~Smolin,
 ``Towards a Theory of Space-Time Structure at Very Short Distances,''
 Nucl. Phys. B \textbf{160} (1979), 253-268
 doi:10.1016/0550-3213(79)90059-2


\bibitem{Kugo}
K.~Hayashi and T.~Kugo,
``Everything about Weyl's gauge field,''
Prog. Theor. Phys. \textbf{61} (1979), 334
doi:10.1143/PTP.61.334

\bibitem{CDA2}
C.~Condeescu and A.~Micu,
``The gauge theory of Weyl group and its interpretation as Weyl quadratic gravity,''
Class. Quant. Grav. \textbf{42} (2025) no.6, 065011
doi:10.1088/1361-6382/adb3e8
[arXiv:2408.07159 [hep-th]].

\bibitem{Frolov}
O.~V.~Babourova and B.~N.~Frolov,
``Pontryagin and Euler forms and Chern-Simons terms in Weyl-Cartan space,''
Mod. Phys. Lett. A \textbf{12} (1997), 1267-1274
doi:10.1142/S0217732397001278
[arXiv:gr-qc/9609005 [gr-qc]].

\bibitem{Jia}
W.~Jia and M.~Karydas,
``Obstruction tensors in Weyl geometry and holographic Weyl anomaly,''
Phys. Rev. D \textbf{104} (2021) no.12, 126031
doi:10.1103/PhysRevD.104.126031
[arXiv:2109.14014 [hep-th]].

\bibitem{ST}
  E.~C.~G.~Stueckelberg,
``Interaction forces in electrodynamics and in the field theory of nuclear forces,''
  Helv.\ Phys.\ Acta {\bf 11} (1938) 299.

\bibitem{LAG}
L.~Alvarez-Gaume, A.~Kehagias, C.~Kounnas, D.~L\"ust and A.~Riotto,
``Aspects of Quadratic Gravity,''
Fortsch. Phys. \textbf{64} (2016) no.2-3, 176-189
doi:10.1002/prop.201500100
[arXiv:1505.07657 [hep-th]].
  
\bibitem{Latorre}
A.~D.~I.~Latorre, G.~J.~Olmo and M.~Ronco,
``Observable traces of non-metricity: new constraints on metric-affine gravity,''
Phys. Lett. B \textbf{780} (2018), 294-299
doi:10.1016/j.physletb.2018.03.002
[arXiv:1709.04249 [hep-th]].


\bibitem{SMW2}
D.~M.~Ghilencea and C.~T.~Hill,
``Standard Model in conformal geometry: Local vs gauged scale invariance,''
Annals Phys. \textbf{460} (2024), 169562
doi:10.1016/j.aop.2023.169562
[arXiv:2303.02515 [hep-th]].

\bibitem{Mannheim2}
  P. D. Mannheim,
  ``Conformal cosmology with no cosmological constant''
  Gen. Rel. Grav. \textbf{22} (1990), 289-298;
 doi:10.1007/BF00756278
  ``Making the Case for Conformal Gravity,''
  Found. Phys. \textbf{42} (2012), 388-420
 doi:10.1007/s10701-011-9608-6
[arXiv:1101.2186 [hep-th]].
 See also
 P. D. Mannheim and J. G. O'Brien,
 ``Fitting the galactic rotation curves with conformal
 gravity and a global quadratic potential'',    Physical Review D {\bf 85}, I 124020 (2012).
 P. D. Mannheim, ``Cosmological perturbations in conformal gravity'',
    Phys. Rev. D  85,  124008 (2012).


\bibitem{int1}
D.~M.~Ghilencea, L.~E.~Ibanez, N.~Irges and F.~Quevedo,
``TeV scale Z-prime bosons from D-branes,''
JHEP \textbf{08} (2002), 016
doi:10.1088/1126-6708/2002/08/016
[arXiv:hep-ph/0205083 [hep-ph]].

\bibitem{int2}
D.~M.~Ghilencea,
``U(1) masses in intersecting $D$-brane SM - like models,''
Nucl. Phys. B \textbf{648} (2003), 215-230
doi:10.1016/S0550-3213(02)00977-X
[arXiv:hep-ph/0208205 [hep-ph]].

\bibitem{planck2018}
  Y.~Akrami {\it et al.} [Planck Collaboration],
  ``Planck 2018 results. X. Constraints on inflation,''
  arXiv:1807.06211 [astro-ph.CO].


\bibitem{CMB1}
  K.~N.~Abazajian {\it et al.} [CMB-S4 Collaboration],
``CMB-S4 Science Book, First Edition,''
  arXiv:1610.02743 [astro-ph.CO].
https://cmb-s4.org/

\bibitem{CMB2}
  J.~Errard, S.~M.~Feeney, H.~V.~Peiris and A.~H.~Jaffe,
``Robust forecasts on fundamental physics from the foreground-obscured, 
gravitationally-lensed CMB polarization,''
  JCAP {\bf 1603} (2016) no.03,  052
  [arXiv:1509.06770 [astro-ph.CO]].


\bibitem{CMB3}
  A.~Suzuki {\it et al.},
``The LiteBIRD Satellite Mission - Sub-Kelvin Instrument,''
  J.\ Low.\ Temp.\ Phys.\  {\bf 193} (2018) no.5-6,  1048
  [arXiv:1801.06987 [astro-ph.IM]].


\bibitem{litebird}
T.~Matsumura et al,
``Mission design of LiteBIRD,''
J. Low Temp. Phys. \textbf{176} (2014), 733
[arXiv:1311.2847 [astro-ph.IM]].

\bibitem{CMB4} 
S.~Hanany \textit{et al.} [NASA PICO],
``PICO: Probe of Inflation and Cosmic Origins,''
[arXiv:1902.10541 [astro-ph.IM]].

\bibitem{Pixie}
A.~Kogut, D.~Fixsen, D.~Chuss, J.~Dotson, E.~Dwek, M.~Halpern, G.~Hinshaw, 
S.~Meyer, S.~Moseley, M.~Seiffert, D.~Spergel and E.~Wollack,
``The Primordial Inflation Explorer (PIXIE): A Nulling Polarimeter
 for Cosmic Microwave Background Observations,''
JCAP \textbf{07} (2011), 025
[arXiv:1105.2044 [astro-ph.CO]].


\bibitem{Misha2}
  R. Armillis, A. Monin, M. Shaposhnikov, “Spontaneously Broken Conformal Symmetry:
Dealing with the Trace Anomaly,” JHEP 1310 (2013) 030 [arXiv:1302.5619 [hep-th]].

\bibitem{Misha2s}
M.~Shaposhnikov and A.~Tokareva,
``Exact quantum conformal symmetry, its spontaneous breakdown, and gravitational Weyl anomaly,''
Phys. Rev. D \textbf{107} (2023) no.6, 065015
doi:10.1103/PhysRevD.107.065015
[arXiv:2212.09770 [hep-th]].

\bibitem{Misha3}
M.~Shaposhnikov and A.~Tokareva,
``Anomaly-free scale symmetry and gravity,''
Phys. Lett. B \textbf{840} (2023), 137898
doi:10.1016/j.physletb.2023.137898
[arXiv:2201.09232 [hep-th]].


\bibitem{Tamarit}
C.~Tamarit,
``Running couplings with a vanishing scale anomaly,''
JHEP \textbf{12} (2013), 098
doi:10.1007/JHEP12(2013)098
[arXiv:1309.0913 [hep-th]].


\bibitem{dg1}
  D.~Ghilencea,
  ``Manifestly scale-invariant regularization and quantum effective operators,''
  Phys.\ Rev.\ D {\bf 93} (2016) no.10,  105006
 doi:10.1103/PhysRevD.93.105006
    [arXiv:1508.00595[hep-ph]].
 ``One-loop potential with scale invariance and effective operators,''
  PoS CORFU {\bf 2015} (2016) 040
    [arXiv:1605.05632 [hep-ph]].


\bibitem{dg2}
  D.~Ghilencea, Z.~Lalak and P.~Olszewski,
  ``Two-loop scale-invariant scalar potential and quantum effective operators,''
  Eur.\ Phys.\ J.\ C {\bf 76} (2016) no.12,  656
  doi:10.1140/epjc/s10052-016-4475-0  [arXiv:1608.05336 [hep-th]].


\bibitem{dg3}
  D.~M.~Ghilencea,
  ``Quantum implications  of a scale invariant regularization,''
  Phys.\ Rev.\ D {\bf 97} (2018) no.7,  075015  doi:10.1103/PhysRevD.97.075015
    [arXiv:1712.06024 [hep-th]].

\bibitem{dg7}
  D.~M.~Ghilencea, Z.~Lalak and P.~Olszewski,
  ``Standard Model with spontaneously broken quantum scale invariance,''
  Phys.\ Rev.\ D {\bf 96} (2017) no.5,  055034
doi:10.1103/PhysRevD.96.055034
  [arXiv:1612.09120 [hep-ph]].

\bibitem{3loop}
  F. Gretsch and A. Monin, ``Perturbative conformal symmetry and dilaton,'' Phys. Rev. D
92 (2015) no.4, 045036 [arXiv:1308.3863 [hep-th]].


\bibitem{Buch}
I. L. Buchbinder, I. L. Shapiro, "Introduction to QFT with applications to Quantum Gravity",
Oxford University Press, Oxford 2021.


\bibitem{Asorey}
M.~Asorey, E.~V.~Gorbar and I.~L.~Shapiro,
``Universality and ambiguities of the conformal anomaly,''
Class. Quant. Grav. \textbf{21} (2003), 163-178
doi:10.1088/0264-9381/21/1/011
[arXiv:hep-th/0307187 [hep-th]].

\bibitem{F1}
P.~G.~Ferreira, C.~T.~Hill and G.~G.~Ross,
``Scale-Independent Inflation and Hierarchy Generation,''
Phys. Lett. B \textbf{763} (2016), 174-178
doi:10.1016/j.physletb.2016.10.036
[arXiv:1603.05983 [hep-th]].

\bibitem{F2}
P.~G.~Ferreira, C.~T.~Hill and G.~G.~Ross,
``No fifth force in a scale invariant universe,''
Phys. Rev. D \textbf{95} (2017) no.6, 064038
doi:10.1103/PhysRevD.95.064038
[arXiv:1612.03157 [gr-qc]].

\bibitem{F3}
P.~G.~Ferreira, C.~T.~Hill and G.~G.~Ross,
``Inertial Spontaneous Symmetry Breaking and Quantum Scale Invariance,''
Phys. Rev. D \textbf{98} (2018) no.11, 116012
doi:10.1103/PhysRevD.98.116012
[arXiv:1801.07676 [hep-th]].

\bibitem{F4}
P.~G.~Ferreira, C.~T.~Hill and G.~G.~Ross,
``Weyl Current, Scale-Invariant Inflation and Planck Scale Generation,''
Phys. Rev. D \textbf{95} (2017) no.4, 043507
doi:10.1103/PhysRevD.95.043507
[arXiv:1610.09243 [hep-th]].

\bibitem{F5}
J.~Garcia-Bellido, J.~Rubio, M.~Shaposhnikov and D.~Zenhausern,
``Higgs-Dilaton Cosmology: From the Early to the Late Universe,''
Phys. Rev. D \textbf{84} (2011), 123504
doi:10.1103/PhysRevD.84.123504
[arXiv:1107.2163 [hep-ph]].

\bibitem{ghosts}
I.~Antoniadis, E.~Dudas and D.~M.~Ghilencea,
``Living with ghosts and their radiative corrections,''
Nucl. Phys. B \textbf{767} (2007), 29-53
doi:10.1016/j.nuclphysb.2006.12.019
[arXiv:hep-th/0608094 [hep-th]].


\bibitem{HH}
S.~W.~Hawking and T.~Hertog,
``Living with ghosts,''
Phys. Rev. D \textbf{65} (2002), 103515
doi:10.1103/PhysRevD.65.103515
[arXiv:hep-th/0107088 [hep-th]].

\bibitem{KS}
K.~S.~Stelle,
``Renormalization of Higher Derivative Quantum Gravity,''
Phys. Rev. D \textbf{16} (1977), 953-969
doi:10.1103/PhysRevD.16.953

\bibitem{DBI}
D.~M.~Ghilencea,
``Weyl gauge invariant DBI action in conformal geometry,''
Phys. Rev. D \textbf{111} (2025) no.8, 085019
doi:10.1103/PhysRevD.111.085019
[arXiv:2407.18173 [hep-th]].

\bibitem{Tasi}
  See for example D.~Tong,
``String Theory,''
[arXiv:0908.0333 [hep-th]].
  
\bibitem{Penrose}
R. Penrose (2006).
``Before the Big Bang: An Outrageous New Perspective and its Implications
for Particle Physics'',  Proceedings of the EPAC 2006, Edinburgh, Scotland:
2759–2762. Bibcode:2006epac.conf.2759R.
V. G. Gurzadyan; R. Penrose (2013).
``On CCC-predicted concentric low-variance circles in the CMB sky''.
Eur. Phys. J. Plus. 128 (2): 22. arXiv:1302.5162.
 doi:10.1140/epjp/i2013-13022-4.
  V.G. Gurzadyan; R. Penrose  (2010-11-16).
 ``Concentric circles in WMAP data may provide evidence of violent pre-Big-Bang activity''
 arXiv:1011.3706 [astro-ph.CO].

\bibitem{Huang}
C.~g.~Huang, D.~d.~Wu and H.~q.~Zheng,
``Cosmological constraints to Weyl's vector meson,''
Commun. Theor. Phys. \textbf{14} (1990), 373-378
BIHEP-TH-89-40.


\bibitem{Tang}
Y.~Tang and Y.~L.~Wu,
``Weyl Symmetry Inspired Inflation and Dark Matter,''
Phys. Lett. B \textbf{803} (2020), 135320
doi:10.1016/j.physletb.2020.135320
[arXiv:1904.04493 [hep-ph]].

}
\end{thebibliography}
\end{document}